\documentclass[showpacs,aps,prb,reprint,twocolumn]{revtex4-1}

\usepackage{commath}
\usepackage{amsmath}
\usepackage{graphicx}
\usepackage{hyperref}
\usepackage{txfonts}

\draft 

\begin{document}


\title{A theory of electrodynamic response for bounded metals: surface capacitive effects} 


\author{Hai-Yao Deng}
\email{haiyao.deng@gmail.com}
\affiliation{School of Physics and Astronomy, Cardiff University, 5 The Parade, Cardiff CF24 3AA, Wales, United Kingdom}

\begin{abstract} 
We report a general macroscopic theory for the electrodynamic response of semi-infinite metals (SIMs). The theory includes the hitherto overlooked capacitive effects due to the finite spatial extension of a surface. The basic structure of this theory is independent of the particulars of electron dynamics. Analytical expressions have been obtained of the charge density-density response function, which is naturally parsed into two parts. One of them represents a bulk property while the other a pure surface property. We apply the theory to study the responses according to several electronic dynamics models and provide a unified view of their validity and limitations. The models studied include the local dielectric model (DM), the dispersive hydrodynamic model (HDM) and specular reflection model (SRM), as well as the less common semi-classical model (SCM) based on Boltzmann's transport equation. We show that, in terms of their basic equations, the SRM is an extension of the HDM, just as the HDM is an extension of the DM. The SCM improves over the SRM critically through the inclusion of translation symmetry breaking and surface roughness effects. We then employ the response function to evaluate the so-called dynamical structure factor, which plays an important role in particle scattering. As expected, this factor reveals a peak due to the excitation of surface plasma waves (SPWs). Surprisingly, however, the peak is shown to be considerably sharper in the SCM than in other models, indicating an incipient instability of the system according to this model. We also study the distribution of charges induced by a charged particle grazing over a SIM surface at constant speed. This distribution is shown to contain model-specific features that are of immediate experimental interest. This work is expected to find broad applications in optics, plasmonics and other areas such as electron energy loss spectroscopy and accelerator design. 
\end{abstract}

\maketitle
\section{introduction}
\label{sec:1}
Electrodynamic responses, that is, the behaviors of charges in materials and the accompanied electromagnetic fields when subjected to external probes, underlie many physical processes involving the interaction of electron and photon with condensed matter. In principle, the response function can be computed with time-dependent density functional theory (TDDFT)~\cite{pitarke2007} or other formalisms such as Greenwood-Kubo theory. In reality, however, the presence of boundaries (i.e. interfaces and surfaces), which exist in any real materials, makes such computation often unrealistic and impractical. Primarily this is because genuine physical boundaries are complicated and their atomistic profiles are unknown \textit{a priori} whereas in microscopic computation they are usually treated in a highly simplified fashion. On numerous occasions, e.g. in studying optical properties, a microscopic boundary is only of marginal importance and a macroscopic description could be more useful. The obvious path to attaining a macroscopic description is to start with a microscopic model and then proceed to the macroscopic limit, resulting in theories that are nevertheless model specific~\cite{chen1993}. 

To establish a macroscopic theory that is as generic as the microscopic one, a major conceptual obstacle needs to be circumvented, which lies with the macroscopic limit of physical boundaries. Let us take for illustration a surface. On the atomistic scale, the surface layer of a material differs from its bulk interior only quantitatively and all microscopic characteristics -- such as the geometric arrangements of atoms and the chemical compositions -- smoothly evolve throughout the system without abrupt changes. On a macroscopic scale, however, the surface layer becomes infinitesimally thin regardless of its microscopic details and it is not at all self-evident how and what general physical effects inherited from the microscopic surface should be dealt with. 

Traditionally, a macroscopic boundary of vanishing thickness has been treated as a geometric separation and the physical quantities on the opposite interior sides of this separation are then related by boundary conditions. Amongst these are the Maxwell's boundary conditions (MBCs), which directly follow from the fundamental equations of electromagnetism and are the basis of the usual rules governing the reflection and transmission of optical rays. For non-dispersive materials, i.e. those whose electric polarization or current density depends locally on the electric field present in them, MBCs suffice for all purposes. However, for dispersive materials MBCs are well known to be insufficient to determine the solutions. To remedy this deficiency, since 1950s additional boundary conditions (ABCs) have been invoked to supplement the MBCs~\cite{pekar1957}. These conditions artificially fix the boundary values of certain physical quantities such as the polarization or current density. Despite their widespread use, ABCs lack universality and experimental support. Efforts of justifying them usually start from some microscopic model and the results are specific to the model in use~\cite{chen1993}. Some authors showed that conditions equivalent to ABCs could be obtained by use of the extinction theorem~\cite{birman1972,wolf1972,schmidt2016,schmidt2018}. They based their results on the so-called 'dielectric approximation', which simply assumed the dispersive constitutive relation of an infinite system extended up to the boundary. This assumption sounds natural but does not take into account genuine boundary effects~\cite{bishop1976}. It is worth noting that none boundary conditions are needed in microscopic approaches. 

In addition to the problem of ABCs, few existing work have discussed the fact that a boundary is not just geometrical but also physical. For example, from a microscopic point of view, the potential in the surface layer differs from the rest and electron waves should be scattered. A macroscopically flat surface can thus appear rough to electron waves that can resolve a distance of the order of a Fermi wavelength. Such scattering effects break translation symmetry and cannot be incorporated in the dielectric approximation. Moreover, however thin it may appear on a macroscopic scale, a surface layer always represents a region in which physical quantities may experience rapid variations. In particular, charges can flow into and out of the layer leading to capacitive effects. A generic way of handling these unique surface effects in macroscopic theories remains to be recognized.  

Our intent here is to put forth a macroscopic theory of electrodynamic response that is applicable to any models of electron dynamics for metals, be they dispersive or non-dispersive. Our theory is based on a straightforward yet general macroscopic description of physical surfaces, which should be valid irrespective of their microscopic profiles, thereby doing away with both MBCs and ABCs just as in the microscopic approaches. This method has recently been used to analyze a simple system (a linear anisotropic dielectric) in a pedagogical way~\cite{deng2020}. With this theory, we analyze the electrodynamic responses according to several widely used dispersive or non-dispersive electron dynamics models. Their limitations and relations as well as some longstanding misconceptions about them are critically reviewed and clarified. We then discuss two experimentally interesting quantities: the dynamical structure factor relevant for particle scattering and the distribution of charges induced by a grazing particle that is relevant for surface absorption profile. Applying the theory to metal screening and the propagation of electromagnetic waves will be presented elsewhere.

In this paper we are interested in high-frequency responses, where the ionic motions can be treated as quasi-static and only electronic motions need to be considered. 

In the rest of this section, we give an overview of existing work (Sec.~\ref{sec:1.1}) and outline our main results (Sec.~\ref{sec:1.2}). 

\subsection{Overview of the literature}
\label{sec:1.1}
Electrodynamic response may be quantified by the charge density-density response function, which measures the amount of charges induced in a material due to certain probing potential. For infinite systems possessing full translation symmetry, this function has been known in details since the work of Bohm and Pines in the 1950s~\cite{bohm,pines,hubbard,eguiluz1995}. Their work established the concept of collective electronic oscillations -- known as plasma waves or more precisely volume plasma waves (VPWs, sometimes called \textit{bulk plasma waves}) -- in the bulk of metals. In reality every system is bounded with surfaces -- the hotbed of novel physics and applications~\cite{duke1994}. For example, shortly after the discovery of VPWs, it was predicted and later experimentally confirmed that similar oscillations could also be sustained on metal surfaces~\cite{ritchie1957,powell1960}. The study of such surface plasma waves (SPWs) has nowadays grown into a vast field called plasmonics~\cite{plasmonics,plasmonics2,plasmonics3}, which has been pitched as the most viable way toward sub-wavelength control of light-matter interaction. VPWs and SPWs typically dominate the response at frequencies much higher than that of lattice vibrations and other low-energy elementary excitations. 

Bounded systems do not possess full translation symmetry and the response function is usually difficult to calculate~\cite{feibelman1982,apell1984,horing1985,tsuei1991,pitarke2007}. Analytical solutions do not generally exist and an adequate generic understanding remains to be achieved properly taking into account the effects of translation symmetry breaking and surface roughness. Existing work are either macroscopic or microscopic~\cite{pitarke2007}, the former based on simple models while the latter relying on computational time-dependent density functional theory. While it accounts for the surface effects in a self-consistent manner and might even provide a microscopic knowledge of the surface itself, the computational approach does not always make transparent the underlying physics and often presumes an ideal surface, such as those modeled by a hard-or-soft-wall-type infinite barrier potential~\cite{kempa1985}. In addition, it can be computationally expensive for studying realistic aspects of surfaces, e.g. roughness~\cite{ullrich2001}. In recent years, there has seen lots of effort to synergize simple models with density functional theory (in the so-called quantum hydrodynamic model~\cite{toscano2015,yan2015,ciraci2016,christensen2017}) so as to take advantage of both approaches. 

Despite the increasing use of computational approaches in electrodynamic response studies~\cite{tsuei1991,pitarke2007,liebsch1993,pitarke2001,silkin2003,silkin2008}, the macroscopic approach with simple models continues to be a useful approach and provide additional insights, in particular in the field of applications. The most widely-used amongst existing models include the local dielectric model (DM)~\cite{ritchie1957,stern1960,kanazawa1961,ritchie1962,ekart1981,babiker1982,abdalaziz1989,pendry1975}, the hydrodynamic model (HDM)~\cite{ritchie1963,bennett1970,harris1971,barton1979,nakamura1983,pendry2013,luo} and the specular reflection model (SRM)~\cite{ritchie1966,flores1979,arista1992,yubero1996}. These models have existed for a long time and they have been frequently employed to understand surface-related phenomena, examples including the surface energy absorption profile~\cite{nunez1980}, the energy loss spectra of particles scattered off metal surfaces~\cite{mills1975,rocca1995,juaristi2005}, the image potential and stopping power~\cite{ekart1981,echenique1981,aminov2001}, the free energy of metals~\cite{chan1975}, van der Waals forces and Casimir forces~\cite{luo}, quantum friction and Coulomb drag between relatively moving objects~\cite{pendry1997,pendry2010,volokitin2011,echenique2011}, ion neutralization spectra~\cite{monreal2014} and energy dissipation and transport in quantum dots in the proximity of metal surfaces~\cite{charles2014,vagov2016,downing2017} as well as photon drag effect (see Ref.~\cite{strait2019} and references therein). The DM presumes a local dependence of the electrical current density on the electric field (Drude's law) and is valid only for non-dispersive medium. Where non-local effects are intended, i.e. in dispersive medium, the HDM and the SRM are usually invoked~\cite{eugiluz1979,gerhardts1983}. 

An immediate issue in dealing with dispersive models is that, the models by themselves are insufficient for determining the responses from a macroscopic point of view. Due to spatial dispersion, knowledge must be supplied of the nature of the surfaces to get a unique solution. Historically, this conceptual deficiency has been remedied by a set of what is now known as auxiliary boundary conditions (ABCs)~\cite{moliner}, which are imposed to fix the surface values of electric currents or polarization. Very commonly, and overwhelmingly in papers working with the HDM~\cite{harris1971,barton1979,pendry2013}, it has been imposed that no normal electrical current flows in the immediate neighborhood of a surface~\cite{pitarke2007,moliner}. In the language of electromagnetism, this translates into the vanishing of electrical polarization on surfaces, a condition that was first introduced by Pekar~\cite{pekar1957,silin1961} in the 1950s and has since been adopted in many variations~\cite{maradudin1973,garcia1977a,garcia1977b,churchill2017}. Nevertheless, it was pointed out long ago that ABCs are physically superficial having no general physical basis~\cite{ginzberg1966,henneberger2010}. Papers trying to justify the ABCs are largely tailored for specific circumstances and lack universality~\cite{forstmann1978,helevi1992,henkel2012,mario2014} or based on extinction theorem type development within the dielectric approximation. In a microscopic description that self-consistently takes care of the surfaces, ABCs are obviously superfluous and in fact experimentally refutable~\cite{tignon2000}. In addition, ABCs are incompatible with the DM and any local models, which are self-sufficient and requires no ABCs. Recently, Henneberger called ABCs \textit{a historical mistake}~\cite{hennenberg1998} and proposed a scheme to remove them. Instead of ABCs, he introduced the concept of a surface acting as a radiation source analogous to the charge sheet in the SRM, which is itself controversial~\cite{nelson1999} and has been refuted by experiments~\cite{tignon2000}.         

Beside the electronic models mentioned above, there is a far less common but more accurate model, namely the semi-classical model (SCM)~\cite{moliner,garcia1977a,garcia1977b}. This model describes electron dynamics by a semi-classical equation of motion and Boltzmann's transport equation. It is perhaps one of the closest to a rigorous quantum mechanical description~\cite{harris1979,aminov2001}. The application of this model in bounded systems dated back to the late 1930s, when Fuchs applied it to study the boundary effects on electric conductivity of thin films~\cite{fuchs1938}. In his work, Fuchs introduced a useful parameter $p$, taking values between zero and unity, to denote the fraction of electrons specularly reflected back off a surface. Circa 1940s, the SCM was used to study anomalous skin effect~\cite{reuter1948,fuchs1968} and has since been developed into the standard theory for dealing with this effect~\cite{ziman,abrikosov,kaganov1997}. In late 1970s, Flores and Garcia were amongst the first to employ it to study electromagnetic responses of dispersive medium on the basis of ABCs~\cite{garcia1977a,garcia1977b}. As an advantage, the SCM allows one to take care of both translation symmetry breaking as well as surface roughness effects, the latter via the \textit{Fuchs} parameter.  

\subsection{Outline of main results}
\label{sec:1.2}
The main purpose of the present work is to derive a macroscopic electrodynamic response theory for semi-infinite metals (SIMs) that is free from the usual boundary conditions, and then employ it to calculate the density-density response function (Sec.~\ref{sec:2}). This is possible thanks to a simple yet general macroscopic description of interfaces possessing whatever microscopic profile. The theory is formulated in a generic form assuming no particulars of electronic dynamics, be they quantum mechanical or classical, local or dispersive. It is valid as long as the length scale of the responses is much bigger than the thickness of the microscopic surface layer so that this layer may be treated as of vanishing thickness, i.e. the macroscopic limit. 

With this theory, it is shown that the response function naturally contains two components, one being essentially the same as for an infinite system whereas the other solely due to the presence of surfaces (Secs.~\ref{sec:2a} and \ref{sec:2b}). It is to the latter that the SPWs contribute. We find that under ABCs the surface contribution would be totally lost and hence no SPWs would exist, in agreement with our recent work~\cite{deng2018} showing that the apparent SPW solution admitted in ABC-based HDM is incompatible with that of the DM. 

The generality of the theory allows to scrutinize the various model-based macroscopic descriptions under one umbrella and disclose their conceptual relations (Sec.~\ref{sec:2c}). Upon inputing a local electron dynamics, the theory expectedly revisits well known results based on the DM. When applied to the classical HDM, our theory yields qualitatively different response due to the surface contribution than the usual treatment of this model based on ABCs imposing no normal current on the surface. 

The SRM is subtle in its original design. Nominally, it assumes a specularly reflecting surface and thus no normal current should flow near the surface in apparent conformity with the usual ABCs. No surface contribution and hence no SPWs should occur in this model. Nonetheless, it additionally assumes the existence of a \textit{fictitious} charge sheet located exactly on the surface. As shown in Ref.~\cite{SI}, this charge sheet partially restores the surface contribution and gives rise to SPWs. As far as SPWs are concerned, the SRM is revisited as a direct extension of the HDM in our theory. Despite this, its two basic assumptions are incompatible and the model does not correspond to the specular reflection limit ($p=1$) of the SCM.  

With the SCM, we thoroughly treat the semi-classical response by the theory in Sec. \ref{sec:3}. The SCM unveils two interesting yet natural features unseen in other models. Firstly, translation symmetry breaking effects drastically modify the surface part of the response function. Secondly, the function shows dependence on surface roughness by virtue of the \textit{Fuchs} parameter $p$. In the specular reflection limit, i.e. $p=1$, the surface contribution disappears and the SRM is not restored, as aforementioned. 

Various defining quantities of the theory and models are summarized in Table~\ref{t:1}, where their relations are made clear. 

In connection with the experimental consequences of the symmetry breaking effects, we discuss briefly the energy loss spectra of charged particles reflected off a metal surface in Sec.~\ref{sec:6}. We calculate the dynamical structure factor within the widely used dipole approximation~\cite{tsuei1991,mills1975,persson1977}. It is found that the SPW peak is asymmetric and exceptionally sharper in the SCM than in other models. Actually, its width can possibly be made to vanish by reducing the thermal electronic collision rate, implying that the system contains an instability. It leads to lossless SPWs at the critical point~\cite{SI}, a highly desirable attribute in plasmonics. This finding defies conventional wisdom~\cite{khurgin2017} but is consistent with empirical facts and agrees with our previous work~\cite{deng2018,deng2017a,deng2017b,deng2017c}, where it was shown that the decay rate of SPWs is not simply a sum of the thermal collision rate, Landau damping rate and other loss rates such as inter-band absorption rates, but should be from these deducted by a positive-definite term, which is guaranteed by the principle of physical causality. 

Another quantity of experimental interest is the spatial distribution of induced charges, which are ultimately responsible for the surface absorption profile and stopping power~\cite{nazarov2007} and may be chartered out directly by near-field-optical microscopy. As an illustration, we have evaluated the distribution of these charges induced by an exterior charged particle grazing over the surface at constant speed [Fig.~\ref{fig:f1} (a)]. The distribution is shown sensitive to which model is in use, see Fig.~\ref{fig:f1} (b) for instance and Sec.~\ref{sec:6} for thorough discussions. For example, according to the DM the induced charges should always be symmetrically deployed about the particle along the direction of its motion, while according to the SCM more charges are concentrated in front of the particle. Perpendicular to the direction of motion, the distribution is periodic in all models but with a much shorter period in the SCM. 

An online supplemental text~\cite{SI} has been provided to discuss various issues that could not be accommodated in the main text. These include a phenomenological model~\cite{liebsch1993,nazarov2016} for partially accounting for the contributions of valence electrons, some numerical details, some properties of SPWs in the SCM and the logical inconsistencies of the conventional SRM. 

\section{Theory of dynamical responses} 
\label{sec:2}
In this section, we derive the macroscopic electrodynamic response theory and calculate the charges induced by external stimuli, and from this the density-density response function is extracted including contributions from both the SPWs and VPWs. The theory is founded on general physical concepts and independent of the particulars of electron dynamics. For the sake of illustration, we shall consider a semi-infinite metal (SIM) with a single macroscopically flat surface. Extension to films and other geometries is straightforward and will be considered elsewhere. 

\begin{figure*}
\begin{center}
\includegraphics[width=0.95\textwidth]{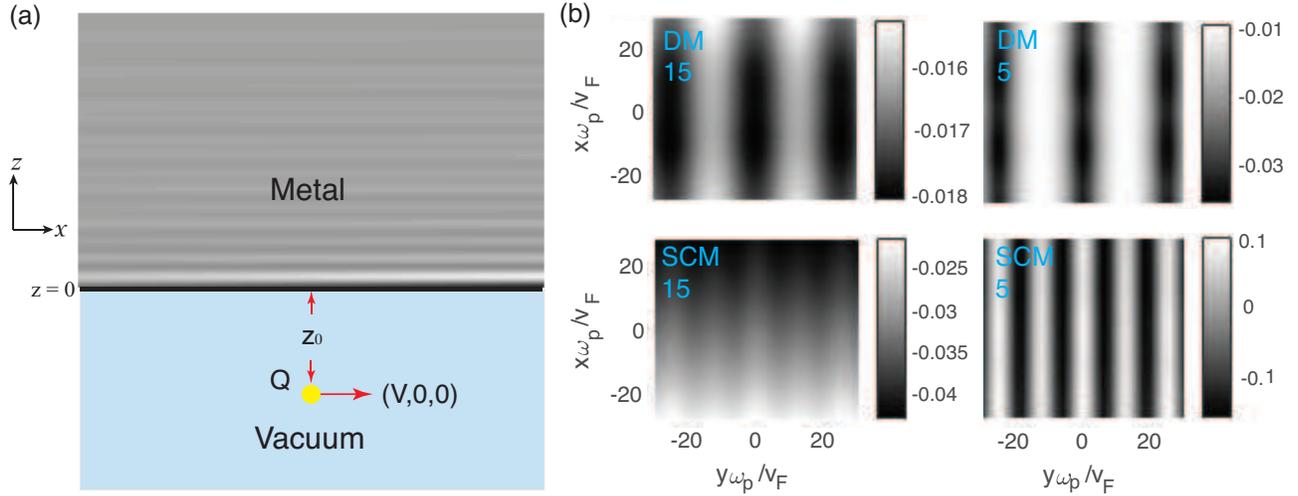}
\end{center}
\caption{(a) Sketch of the system: a semi-infinite metal (SIM) occupies the half space $z\geq0$ and the vacuum occupies the other half. $\mathbf{x} = (\mathbf{r},z)$ and $\mathbf{r} = (x,y)$. A point on the surface is denoted by $\mathbf{x}_0 = (\mathbf{r},0)$. The present work is devoted to deriving a general dynamical response theory for the SIM without suffering from the routinely used boundary conditions. The theory allows us to calculate the charge density $\rho(\mathbf{x},t)$ induced in the SIM due to the presence of any stimuli. In the example shown in panel (a), a particle of unit charge -- indicated by a yellow dot -- grazes over the surface at distance $z_0$ and constant velocity $\mathbf{V} = (V,0,0)$, where $V=10v_F$ for the plot. The gray scale indicates the value of $\rho(\mathbf{x},t)$ in this example. The planar charge distribution, i.e. $\rho_\parallel(\mathbf{r},t) = \int dz~\rho(\mathbf{x},t)$ is displayed in (b) for two models, the DM and the SCM, see Sec.~\ref{sec:6} for discussions and other models. The particle is located at $(0,0,-z_0)$ for the moment under consideration. The number in each panel indicates the value of $z_0\omega_p/v_F$. \label{fig:f1}}
\end{figure*} 

The SIM is assumed to occupy the half-space $z\geq0$ and interfaces with the vacuum at $z=0$, as shown in Fig.~\ref{fig:f1}. Throughout the paper, we reserve $\mathbf{r} = (x,y)$ for planar coordinates and $\mathbf{x} = (\mathbf{r},z)$. A point on the surface is denoted by $\mathbf{x}_0 = (\mathbf{r},0)$ and we use $t$ to denote time. The surface may appear rough on the scale of Fermi wavelength and hence cause diffuse scattering of electron waves, but is assumed sufficiently flat on a macroscopic length scale so that the translational symmetry along the surface is preserved. 

In studying dynamical responses for bounded medium, it is customary to work directly with the electrostatic potential -- or more generally the electromagnetic field in the case of non-negligible retardation effects -- and write down its expressions on the vacuum side and the material side separately. ABCs are then invoked together with the usual MBCs -- the continuity of both the electrostatic potential and the normal component of the electric displacement field in the electrostatic limit -- to join them at the boundary. In what follows we show how a general response theory can be derived without the use of any explicit boundary conditions and other type of \textit{ad hoc} prescriptions such as those of Ref.~\cite{hennenberg1998}. To this end, we first need to establish the macroscopic limit of an arbitrary physical interface in a general way. Considering that a real microscopic surface can hardly be specified even for the simplest material, one might deem it hopeless. However, the following elementary analysis suggests otherwise.  

\begin{figure*}
\begin{center}
\includegraphics[width=0.95\textwidth]{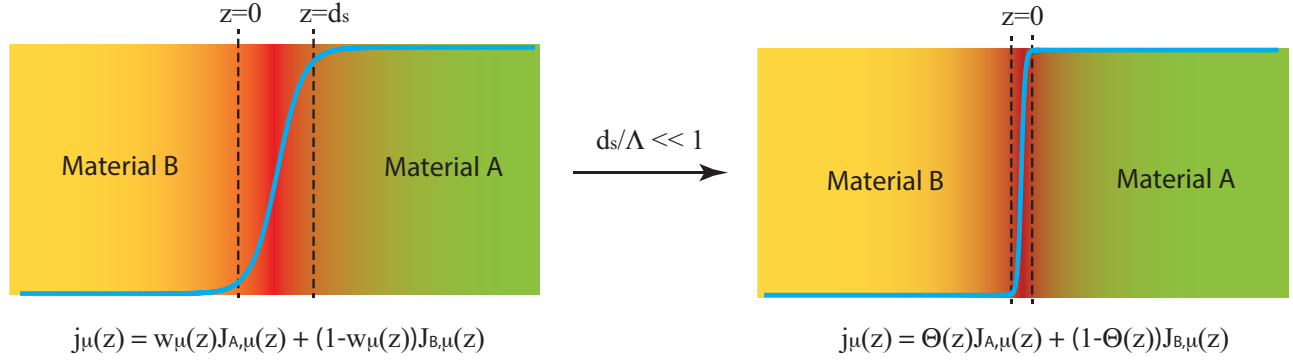}
\end{center}
\caption{The macroscopic limit of a physical interface joining materials A and B. On the atomistic scale, the interface has finite thickness $d_s$ (left). The current density $j_\mu$ can be related to its values $J_{A/B,\mu}$ in the bulk regions (outside the interfacial layer) via the profile functions $w_\mu(z)$, which approaches unity on side A and zero on side B. On a macroscopic length scale $\Lambda\gg d_s$, the interfacial layer appears infinitely thin and $w_\mu(z)$ reduces to Heaviside step function $\Theta(z)$ (right). \label{fig:interface}}
\end{figure*}

Let us imagine bringing two materials (A and B) in contact, and an interfacial layer of thickness $d_s$ -- in the order of a few lattice constants -- shall form in between (see Fig.~\ref{fig:interface}). We may characterize this layer by a surface potential $\phi_s$, which should quickly decay to zero in the bulk regions outside the interfacial layer. The exact microscopic profile of the layer varies from one case to another and can hardly be known \textit{a priori}. Despite this, we may still write down a generic form for the electric current density $\mathbf{j}(\mathbf{x},t)$ in the whole system including the interfacial layer. To this end, we observe that in the bulk regions where $\phi_s$ vanishes, the form of $\mathbf{j}(\mathbf{x},t)$ can be completely determined with the respective dynamic equations for the infinite materials, apart from some parameters (such as the \textit{Fuchs} parameter, see Sec.~\ref{sec:3}) that encode the effects of surface scattering on the electron waves. Let us denote by $\mathbf{J}_{A/B}(\mathbf{x},t)$ the values of $\mathbf{j}(\mathbf{x},t)$ in the bulk region of A/B. Microscopically, $\mathbf{j}$ evolves from $\mathbf{J}_A$ in the bulk region of A, through a rapid but smooth variation in the interfacial layer, to $\mathbf{J}_B$ in the bulk region of B. Formally, we can write for the $\mu$-th component of the current density as $$j_\mu(\mathbf{x},t) = J_{A,\mu}(\mathbf{x},t) w_\mu(z) + J_{B,\mu}(\mathbf{x},t)(1-w_\alpha(z)),$$ where the profile functions $w_\alpha(z)$ approach unity in the bulk region of A and zero in that of B. The exact form of $w_\mu(z)$ depends on the microscopic details of the interfacial layer. On the macroscopic length scale of $\Lambda$, however, the interfacial layer appears infinitely thin and $w_\alpha(z)$ reduce to the Heaviside step function $\Theta(z)$, where $\Theta(z\geq0) = 1$ and $\Theta(z<0)=0$. In the macroscopic limit, one thus ends up with 
\begin{equation}
\mathbf{j}(\mathbf{x},t) = \mathbf{J}_A(\mathbf{x},t)\Theta(z) + \mathbf{J}_B(\mathbf{x},t)(1-\Theta(z)), \label{sur}
\end{equation}
which holds valid for any $w_\mu(z)$ and is thus a general and complete macroscopic description of a physical interface, as long as the perturbation on one side does not cause significant responses on the other~\cite{deng2020b}. 

To recapitulate, Eq.~(\ref{sur}) elegantly captures two important physical consequences of an interface: the rapid variation of the current density through the step function $\Theta(z)$ and the surface scattering effects on electron dynamics through the parameters contained in the bulk values $\mathbf{J}_{A/B}$. These scattering effects -- including the symmetry breaking effects -- have been ignored in most models except for the SCM. In general $\mathbf{J}_A$ and $\mathbf{J}_B$ are not equal on the interface, as is certainly the case for local dynamics models, and charges can then accumulate in the interfacial layer. Such capacitive effects would be mistakenly erased under usual ABCs, which often dictate continuity of current density across an interface, e.g. the vanishing of the normal component at the metal-vacuum interface.

\subsection{Generic formulation}
\label{sec:2a} 
With the macroscopic limit of physical interfaces, Eq.~(\ref{sur}), we now formulate a general theory of electrodynamic response for the SIM. 

We started with the fact that, in response to a probing electric field $\mathbf{E}_\text{probe}(\mathbf{x},t)$ an electrical current flows in the metal and charges may appear, whose density we denote by $\rho(\mathbf{x},t)$. In the jellium model adopted here, $\rho(\mathbf{x},t) = en(\mathbf{x},t)$ is simply the \textit{deviation} $n(\mathbf{x},t)$ of the density of electrons from its mean value $n_0$. Here $e$ is the charge of an electron. These charges then generate an additional electric field denoted by $\mathbf{E}(\mathbf{x},t)$. In the bulk region of the metal, the total electric field felt by the electrons is $\mathbf{E}_\text{tot}(\mathbf{x},t) = \mathbf{E}_\text{probe}(\mathbf{x},t) + \mathbf{E}(\mathbf{x},t)$. In the regime of linear responses, the density of the current flowing in that region then contains two parts, $\mathbf{J}_\text{tot}(\mathbf{x},t) = \mathbf{J}_\text{probe}(\mathbf{x},t) + \mathbf{J}(\mathbf{x},t)$, where $\mathbf{J}_\text{probe}$ and $\mathbf{J}$ are due to $\mathbf{E}_\text{probe}$ and $\mathbf{E}$, respectively. It should be clear that here the responses (and later the conductivity) are defined with respect to the system's Hamiltonian excluding the long-range Coulomb interaction, which is treated by a self-consistent mean field. This can be justified in the random phase approximation or by the TDDFT.  

According to Eq.~(\ref{sur}), the current density throughout the entire space including the vacuum can then be written as 
\begin{equation}
\mathbf{j}(\mathbf{x},t) = \Theta(z)\mathbf{J}_\text{tot}(\mathbf{x},t), \label{gj}
\end{equation}
This relation is implicit in any local dielectric models~\cite{chen}. As to be seen, boundary conditions, i.e. both MBCs and ABCs, are no longer needed. By Eq.~(\ref{gj}) charges can accumulate on the surface layer producing capacitive effects, which would be mistakenly excluded under ABCs or other equivalent prescriptions such as in the hard-wall picture often adopted in computational approaches. The need to go beyond the hard-wall picture has recently drawn considerable attention in the computational hydrodynamic approach~\cite{toscano2015,yan2015,ciraci2016,christensen2017} in studying the local plasmon resonances on metal nano-particles.  

Now the equation of continuity can be used to relate $\rho$ and $\mathbf{j}$ as follows
\begin{equation}
\mathcal{D}_t\rho(\mathbf{x},t) + \partial_\mathbf{x}\cdot\mathbf{j}(\mathbf{x},t) = 0, \quad \mathcal{D}_t =  \tau^{-1}+\partial_t. \label{2a1}
\end{equation} 
Here a global relaxation term $-\rho(\mathbf{x},t)/\tau$ has been included to account for the relaxation of local non-equilibrium charges [due to finite density deviation $n(\mathbf{x},t)$] due to microscopic electronic collisions driving the system toward thermodynamic equilibrium~\cite{aminov2001,deng2017c} [in which the deviation $n(\mathbf{x},t)$ must vanish]. In terms of $\mathbf{J}_\text{tot}$, the equation reads
\begin{equation}
\mathcal{D}_t\rho(\mathbf{x},t) + \partial_\mathbf{x}\cdot\mathbf{J}_\text{tot}(\mathbf{x},t) = -\Theta'(z)J_\text{tot,z}(\mathbf{x}_0,t), \label{2a2}
\end{equation}
where $\Theta'(z)=d\Theta(z)/dz$. In this equation, we have dropped $\Theta(z)$ on the left hand side to simplify the notation, as is clear that $\mathbf{x}$ represents a point on the metal side. To avoid ambiguity, $\Theta'(z)$ should not be simply identified with the Dirac function $\delta(z)$, because $\int^\infty_0dz\Theta'(z) = 1$ by definition but $\int^\infty_0dz\delta(z) = \frac{1}{2}$. The right-hand term of Eq.~(\ref{2a2}) corresponds to the aforementioned capacitive effects. It plays a critical role in the energy conversion process, which has been overlooked until our recent work~\cite{deng2017c}. This term was noticed by A. L. Fetter in his study of edge plasmon in confined two-dimension electron gases~\cite{fetter1980} and also used in Refs.~\cite{vaman2009} in a different context. 

For studying responses, it is convenient to isolate the terms due to the probing field. Thus, we rewrite Eq.~(\ref{2a2}) as 
\begin{equation}
\mathcal{D}^2_t\rho(\mathbf{x},t) + \mathcal{D}_t \partial_\mathbf{x}\cdot \mathbf{J}(\mathbf{x},t) = S(\mathbf{x}_0,t) + S_\text{probe}(\mathbf{x},t), \label{2a3}
\end{equation}
where $S(\mathbf{x}_0,t) = -\Theta'(z) \mathcal{D}_t J_z(\mathbf{x}_0,t)$ and 
\begin{equation}
S_\text{probe}(\mathbf{x},t) = -\mathcal{D}_t\left[\partial_\mathbf{x}\cdot \mathbf{J}_\text{probe}(\mathbf{x},t)+\Theta'(z)J_\text{probe,z}(\mathbf{x}_0,t)\right] \label{2a4}
\end{equation}
denotes the probing source. Introducing the following Fourier transform 
\begin{equation}
\rho(\mathbf{x},t) = \sum_\mathbf{k} \int^\infty_{-\infty}\frac{d\omega}{2\pi}\frac{e^{i(\mathbf{k}\cdot\mathbf{r}-\omega t)}}{\sqrt{A}} \rho(z;\mathbf{k},\omega), \label{2a5}
\end{equation}
where $A$ is the surface area used to quantize the in-plane wave vector $\mathbf{k}$, for the charge density, and analogously for all other fields, we can rewrite Eq.~(\ref{2a3}) as
\begin{equation}
-i\bar{\omega}\nabla\cdot\mathbf{J}(z;\mathbf{k},\omega) - \bar{\omega}^2\rho(z;\mathbf{k},\omega) = S(\mathbf{k},\omega)\Theta'(z) + S_\text{probe}(z;\mathbf{k},\omega). \label{2a6}
\end{equation} 
Here $\bar{\omega} = \omega + i/\tau$, $\nabla = (i\mathbf{k},\partial_z)$ and $S(\mathbf{k},\omega) = i\bar{\omega}J_z(0;\mathbf{k},\omega)$ does not depend on $z$. In the regime of linear responses considered throughout this paper, we can write $\mathbf{J}(z;\mathbf{k},\omega)$ as a linear functional of $\mathbf{E}(z;\mathbf{k},\omega)$, i.e.
\begin{equation}
J_\mu(z;\mathbf{k},\omega) = \sum_{\nu=x,y,z}\int dz' \sigma_{\mu\nu}(z,z';\mathbf{k},\omega) E_\nu(z';\mathbf{k},\omega),
\end{equation}
where $\sigma_{\mu\nu}(z,z';\mathbf{k},\omega)$ is the conductivity tensor by definition. The same relation holds valid between $\mathbf{J}_\text{probe}(z;\mathbf{k},\omega)$ and $\mathbf{E}_\text{probe}(z;\mathbf{k},\omega)$. Now that $\mathbf{E}(z;\mathbf{k},\omega)$ is also a linear functional of $\rho(z;\mathbf{k},\omega)$ by the laws of electrostatics, we can always define a linear operator $\mathcal{H}$ so that
\begin{equation}
\mathcal{H}\rho(z;\mathbf{k},\omega) = -i\bar{\omega}\nabla\cdot\mathbf{J}(z;\mathbf{k},\omega). \label{2a8}
\end{equation}
With this Eq.~(\ref{2a6}) becomes
\begin{equation}
\left(\mathcal{H}-\bar{\omega}^2\right)\rho(z;\mathbf{k},\omega) = S(\mathbf{k},\omega)\Theta'(z) + S_\text{probe}(z;\mathbf{k},\omega). \label{2a9}
\end{equation}
We can do some further transformations by noting that for any quantity existing in the half space a cosine Fourier transform can be defined, i.e.
\begin{equation}
\rho(z;\mathbf{k},\omega) = \frac{2}{\pi}\int^\infty_0 dq ~ \cos(qz) ~ \rho(\mathbf{K},\omega), \label{2a10}
\end{equation}
where $\mathbf{K} = (\mathbf{k},q)$. In terms of $\rho(\mathbf{K},\omega)$, Eq.~(\ref{2a9}) is rewritten as
\begin{eqnarray}
&~&\int^\infty_0 dq' \left\{\mathcal{H}(q,q';\mathbf{k},\omega)-\bar{\omega}^2\delta(q-q')\right\}\rho(\mathbf{K}',\omega) \nonumber\\ &~& \quad \quad \quad \quad \quad \quad \quad \quad \quad \quad  = S(\mathbf{k},\omega)+S_\text{probe}(\mathbf{K},\omega), \label{2a11}
\end{eqnarray}
where $\mathbf{K}'=(\mathbf{k},q')$, $\mathcal{H}(q,q';\mathbf{k},\omega)$ is the matrix element between the cosine waves $\cos(qz)$ and $\cos(q'z)$, and
\begin{equation}
S_\text{probe}(\mathbf{K},\omega) = \int^\infty_0dz ~ \cos(qz) ~ S_\text{probe}(z;\mathbf{k},\omega). \label{2a12}
\end{equation}
To close Eq.~(\ref{2a11}), we utilize the fact that $J_z(0;\mathbf{k},\omega)$ and hence $S(\mathbf{k},\omega)$ are also linear functionals of the charge density, i.e. \begin{equation}
S(\mathbf{k},\omega) = \int^\infty_0 dq ~ \frac{G(\mathbf{K},\omega)}{K^2} ~ \rho(\mathbf{K},\omega) \label{2a13}
\end{equation}
with $K^2 = k^2+q^2$ and $k = \abs{\mathbf{k}}$. Here $G(\mathbf{K},\omega)/K^2$ denotes the kernel, which is material and model specific; see what follows. 

Equations (\ref{2a11}) and (\ref{2a13}) comprise a complete dynamical response theory for SIMs, allowing us to determine the induced charges provided $S_\text{probe}(\mathbf{K},\omega)$ is known. No boundary conditions have been explicitly invoked in this theory. Extension to other geometries such as films and spherical particles will be performed in a separate publication. 

\subsection{Induced charge densities}
\label{sec:22}
Here we obtain the induced charge densities from the theory derived above. 

It is not necessary but useful to simplify the equations in the first place. We can make use of some general properties of $\mathcal{H}(q,q';\mathbf{k},\omega)$ to this end. It is instructive to look at the equations for self-sustained waves in the absence of probing fields, i.e. we leave out $S_\text{probe}$ from Eq.~(\ref{2a11}). As shown in Refs.~\cite{deng2018,deng2017a,deng2017b}, the resulting equation admits of two types of solutions representing VPWs and SPWs, respectively. Those of VPWs satisfy $S(\mathbf{k},\omega) \equiv 0$, and then the VPW frequencies are obtained as solutions to the secular equation $\abs{\mathcal{H}-\bar{\omega}^2} = 0$. As such, we see that $\mathcal{H}$ contains complete information of VPWs in a SIM. It is reasonable to assume that VPWs are not sensitive to the presence of boundaries, and $\mathcal{H}$ is essentially that of an infinite system. To make this statement accurate, let us analyze the conductivity tensor $\sigma_{\mu\nu}(z,z';\mathbf{k},\omega)$, which contains all information of the electron dynamics of the underlying material. For an infinite system without the surface, the translational symmetry is also preserved along $z$-axis and thus $\sigma_{\mu\nu}(z,z';\mathbf{k},\omega)$ depends only on the difference between $z$ and $z'$. However, for a SIM, the symmetry is broken and it must depend on the coordinates individually. It is then useful to decompose $\sigma_{\mu\nu}$ into two parts, $\sigma_{b,\mu\nu}(z-z';\mathbf{k},\omega)$ and $\sigma_{s,\mu\nu}(z,z';\mathbf{k},\omega)$, where $\sigma_{b,\mu\nu}(z-z';\mathbf{k},\omega)$ is that of the infinite system while $\sigma_{s,\mu\nu}(z,z';\mathbf{k},\omega)$ signifies symmetry breaking effects. By Eq.~(\ref{2a8}), $\mathcal{H}$ accordingly splits into two parts, $\mathcal{H}_b$ and $\mathcal{H}_s$. Since it is responsible for the properties of VPWs in an infinite (isotropic) system, $\mathcal{H}_b$ must be diagonal in the $q$-space, i.e. $\mathcal{H}_b(q,q';\mathbf{k},\omega) = \Omega^2(\mathbf{K},\omega)\delta(q-q')$, where $\Omega(\mathbf{K},\omega)$ is a frequency. By virtue of the rotational symmetry of an infinite system, $\Omega$ depends on the length but not the direction of $\mathbf{K}$. In the meanwhile, $\mathcal{H}_s$ gives rise to scattering of VPWs, which generally makes a small perturbation of the order $kv_F/\omega_p$, where $v_F$ is the Fermi velocity of the metal and $\omega_p$ is the characteristic plasma frequency (see the next subsection), and can be treated perturbatively~\cite{deng2017a,deng2017b,deng2017c}. To the zero-th order in this perturbation, we have
\begin{equation}
\mathcal{H}(q,q';\mathbf{k},\omega) = \Omega^2(K, \omega)\delta(q-q'). \label{2a14}
\end{equation}
Equation (\ref{2a11}) then becomes
\begin{equation}
\left[\Omega^2(K,\omega) - \bar{\omega}^2\right]\rho(\mathbf{K},\omega) = S(\mathbf{k},\omega) + S_\text{probe}(\mathbf{K},\omega). \label{2a15}
\end{equation}
It is easy to show that the dielectric function of an infinite system is given by
\begin{equation}
\epsilon(K,\omega) = 1 - \frac{\Omega^2(K,\omega)}{\bar{\omega}^2}. \label{2a16}
\end{equation}
As usual, the zeros of $\epsilon(K,\omega)$ yield the VPW frequencies. The positive-definite quantity $-\text{Im}\left[\epsilon^{-1}(K,\omega)\right]$ is the so-called loss function for an infinite system. Here Im/Re$[f]$ takes the imaginary/real part of an arbitrary quantity $f$. 

Analogously, we may split $G$, the kernel in Eq.~(\ref{2a13}), into two parts, $G_b$ and $G_s$, which originate from $\sigma_{b,\mu\nu}$ and $\sigma_{s,\mu\nu}$, respectively. In all the models to be discussed in this paper, we find that $G_b = -4i\bar{\omega}k\sigma(\omega)$ independent of $q$, where $\sigma(\omega)$ is the local part of $\sigma_{b,\mu\nu}$, namely $\delta_{\mu\nu}\delta(z-z')\sigma(\omega)$ with $\delta_{\mu\nu}$ being the Kronecker symbol. If inter-band transitions are neglected, one further finds $\sigma(\omega) = (i/\bar{\omega})(\omega^2_p/4\pi)$, which is the Drude conductivity. Thus, we arrive at
\begin{equation}
G(\mathbf{K},\omega) = (k/\pi)\omega^2_p + G_s(\mathbf{K},\omega). \label{2aG}
\end{equation}
As to be seen later, in all the models discussed in this paper, except for the SCM, $G_s$ vanishes. 

Combining Eqs.~(\ref{2a13}) and (\ref{2a15}), we easily obtain the density of the induced charges in two components, $$\rho(\mathbf{K},\omega) = \rho_1(\mathbf{K},\omega) + \rho_2(\mathbf{K},\omega),$$ where $\rho_1$ stems directly from $S_\text{probe}$ by Eq.~(\ref{2a15}), i.e. 
\begin{equation}
\rho_1(\mathbf{K},\omega) = \frac{S_\text{probe}(\mathbf{K},\omega)}{\Omega^2(K,\omega) - \bar{\omega}^2} = -\frac{S_\text{probe}(\mathbf{K},\omega)}{\epsilon(K,\omega)}\frac{1}{\bar{\omega}^2}, \label{2a17}
\end{equation}
and $\rho_2$ originates from $S$, which would have been erroneously left out had we imposed that $J_\text{tot,z}(0;\mathbf{k},\omega)\equiv 0$ or other ABCs. This part is given by
\begin{equation}
\rho_2(\mathbf{K},\omega) = -\frac{S(\mathbf{k},\omega)}{\epsilon(K,\omega)}\frac{1}{\bar{\omega}^2} = -\frac{\bar{S}_\text{probe}(\mathbf{k},\omega)}{\epsilon_s(\mathbf{k},\omega)\epsilon(K,\omega)}\frac{1}{\bar{\omega}^2}. \label{2a18}
\end{equation}
Here $\epsilon_s$ and $\bar{S}_\text{probe}$ are defined as
\begin{equation}
\epsilon_s(\mathbf{k},\omega) = 1 + \int^\infty_0 dq' \frac{G(\mathbf{K}',\omega)}{\bar{\omega}^2K^{'2}}\frac{1}{\epsilon(K',\omega)}, \label{2a19}
\end{equation}
which may be called the surface dielectric function, and
\begin{equation}
\bar{S}_\text{probe}(\mathbf{k},\omega) = -\int^\infty_0 dq' \frac{G(\mathbf{K}',\omega)}{\bar{\omega}^2K^{'2}} \frac{S_\text{probe}(\mathbf{K}',\omega)}{\epsilon(K',\omega)}. \label{2a20}
\end{equation} 
By virtue of the rotational symmetry about $z$-axis, we expect that $\epsilon_s(\mathbf{k},\omega)$ depends on the length of $\mathbf{k}$ but not its direction. 

Obviously, $\rho_1(\mathbf{K},\omega)$ features a resonance near the zeros of $\epsilon(K,\omega)$, indicating the excitation of VPWs. On the other hand, $\rho_2(\mathbf{K},\omega)$ contains an additional resonance near the zeros of $\epsilon_s(k,\omega)$. This resonance corresponds to the excitation of SPWs. As shown in Refs.~\cite{deng2018,deng2017a,deng2017b}, the SPW dispersion relation is determined by the equation that $\epsilon_s(k,\omega) = 0$. Further discussions of this equation and the properties of SPWs are presented in Ref.~\cite{SI}. Unlike $\epsilon^{-1}(K,\omega)$, the imaginary part of $\epsilon^{-1}_s(k,\omega)$ does not keep a single sign in the entire spectrum of $\omega \geq 0$. As to be seen later, in the vicinity of VPW resonances there is nearly complete cancellation between the responses encoded in $\rho_1$ and $\rho_2$ under certain circumstances, leaving only the resonance of the SPWs discernible. 

For the sake of completeness, let us also give the electric field generated by the induced charges. The electrostatic potential $\phi(z;\mathbf{k},\omega)$ is given by
\begin{equation}
\phi(z;\mathbf{k},\omega) = \frac{2\pi}{k} \int^\infty_{-\infty} dz' e^{-k\abs{z-z'}}\rho(z';\mathbf{k},\omega). \label{2a22} 
\end{equation}
The electric field is obtained as $\mathbf{E}(z;\mathbf{k},\omega) = -\nabla \phi(z;\mathbf{k},\omega)$. Explicitly, one finds \textit{in the metal} the projection onto the surface
\begin{equation}
    \mathbf{E}_\parallel (z;\mathbf{k},\omega) = -i\int^\infty_0dq ~ \frac{4 \mathbf{k} \rho(\mathbf{K},\omega)}{K^2}\left(2\cos(qz) - e^{-kz}\right) \label{ex}
\end{equation}
and the normal component
\begin{equation}
    E_z(z;\mathbf{k},\omega) = \int^\infty_0dq ~ \frac{4k \rho(\mathbf{K},\omega)}{K^2}\left(2\frac{q}{k}\sin(qz) - e^{-kz}\right). \label{ez}
\end{equation}
These expressions are easily established from the laws of electrostatics. 

\subsection{The density-density response function}
\label{sec:2b}
In this subsection, we discuss two cases of special importance in many applications such as particle and light scattering. The density-density response function will be obtained.  

\textbf{Case (i).} We place some charges exterior to the metal and look at the responses of the metal to these charges. Let the density of these charges be $\rho_\text{ext}(z;\mathbf{k},\omega)$, which exists only on the vacuum side $z<0$. The probing field is obtained from the corresponding electrostatic potential $\phi_\text{probe}(z;\mathbf{k},\omega)$ in the metal. Adapting Eq.~(\ref{2a22}) to this case, we find 
\begin{equation}
\phi_\text{probe}(z\geq0;\mathbf{k},\omega) = \left(e^{-kz}/k\right) \xi(\mathbf{k},\omega), \label{2a25}
\end{equation}
where $$\xi(\mathbf{k},\omega) = 2\pi \int^0_{-\infty} dz ~ e^{kz} \rho_\text{ext}(z;\mathbf{k},\omega).$$ It follows that in the metal
\begin{equation}
\mathbf{E}_\text{probe}(z;\mathbf{k},\omega) = -\nabla \phi_\text{probe}(z;\mathbf{k},\omega) = \xi(\mathbf{k},\omega)e^{-kz}(-i\hat{\mathbf{k}},1). \label{2a26}
\end{equation}
Here $\hat{\mathbf{k}} = \mathbf{k}/k$. Note that this field cannot be used to unveil the complete $q$-resolved profile of the density response of SIMs, as it has a fixed $z$-dependence of the form $e^{-kz}$, regardless of the configuration of the exterior charges. 

The resulting $S_\text{probe}$ is proportional to $\xi$. We can write it as 
\begin{equation}
S_\text{probe}(\mathbf{K},\omega) = B(\mathbf{K},\omega)\xi(\mathbf{k},\omega), \label{2a27}
\end{equation}
where $B(\mathbf{K},\omega)$ is the coefficient. From Eqs.~(\ref{2a17}) and (\ref{2a18}) one finds
\begin{equation}
\rho(\mathbf{K},\omega) = P(\mathbf{K},\omega)\xi(\mathbf{k},\omega), \label{2a28}
\end{equation}
where $P = P_1 + P_2$, with
\begin{equation}
P_1(\mathbf{K},\omega) = -\frac{B(\mathbf{K},\omega)}{\epsilon(K,\omega)}\frac{1}{\bar{\omega}^2}
\end{equation}
and
\begin{equation}
P_2(\mathbf{K},\omega) = -\frac{\bar{B}(\mathbf{k},\omega)}{\epsilon_s(\mathbf{k},\omega)\epsilon(K,\omega)}\frac{1}{\bar{\omega}^2}. \label{2a30}
\end{equation}
Here 
\begin{equation}
\bar{B}(\mathbf{k},\omega) = -\int^\infty_0 dq' \frac{G(\mathbf{K}',\omega)}{\bar{\omega}^2K^{'2}} \frac{B(\mathbf{K}',\omega)}{\epsilon(K',\omega)}. \label{2a31}
\end{equation}
Note that $B(\mathbf{K},\omega)$ depends on the model of electron dynamics. 

\textbf{Case (ii).} We place the metal in an electrostatic potential of the form $$\phi_\text{probe}(z;\mathbf{k},\omega) = \varphi(\mathbf{K}',\omega)\cos(q'z)$$ with $q'$ fixed. The corresponding probing field is given by
\begin{equation}
\mathbf{E}_\text{probe}(z;\mathbf{k},\omega) = \varphi(\mathbf{K}',\omega)\left(-i\mathbf{k}\cos(q'z),q'\sin(q'z)\right). \label{2a32}
\end{equation}
This field implies a probing charge of density $$\rho_\text{probe}(z;\mathbf{k},\omega) = (K^{'2}/4\pi)\varphi(\mathbf{K}',\omega)\cos(q'z),$$ or equivalently $$\rho_\text{probe}(\mathbf{K},\omega) = (K^{'2}/8)\varphi(\mathbf{K}',\omega)\delta(q-q'),$$ which allows us to unveil the $q$-resolved density responses of a SIM. 

Now $S_\text{probe}$ is proportional to $\varphi(\mathbf{K}',\omega)$, i.e.
\begin{equation}
S_\text{probe}(\mathbf{K},\omega) = C(\mathbf{K},\mathbf{K}',\omega)\varphi(\mathbf{K}',\omega),
\end{equation}
where $C(\mathbf{K},\mathbf{K}',\omega)$ is a model-specific coefficient depending on both $q$ and $q'$. The density of the induced charges can now be written as
\begin{equation}
\rho(\mathbf{K},\omega) = \chi(\mathbf{K},\mathbf{K}',\omega)\varphi(\mathbf{K}',\omega). \label{2a34} 
\end{equation} 
Of course, $\chi(\mathbf{K},\mathbf{K}',\omega)$ is nothing but the charge density-density response function for a SIM, which is usually studied with the Greenwood-Kubo formalism. It can be parsed as $\chi = \chi_1+\chi_2$, with
\begin{equation}
\chi_1(\mathbf{K},\mathbf{K}',\omega) = -\frac{C(\mathbf{K},\mathbf{K}',\omega)}{\epsilon(K,\omega)}\frac{1}{\bar{\omega}^2} \label{2a35}
\end{equation}
and
\begin{equation}
\chi_2(\mathbf{K},\mathbf{K}',\omega) = -\frac{\bar{C}(\mathbf{K}',\omega)}{\epsilon_s(\mathbf{k},\omega)\epsilon(K,\omega)}\frac{1}{\bar{\omega}^2}. \label{2a36}
\end{equation}
Here 
\begin{equation}
\bar{C}(\mathbf{K}',\omega) = -\int^\infty_0 dq \frac{G(\mathbf{K},\omega)}{\bar{\omega}^2K^{2}} \frac{C(\mathbf{K},\mathbf{K}',\omega)}{\epsilon(K,\omega)}. \label{2a37}
\end{equation}
The response function in real space, given by 
\begin{eqnarray}
&~& \chi(z,z';\mathbf{k},\omega) = \left(\frac{2}{\pi}\right)^2 \\ &~& \quad \times\int^\infty_0dq'\int^\infty_0dq \cos(q'z') \chi(\mathbf{K},\mathbf{K}',\omega) \cos(qz), \nonumber
\end{eqnarray}
is more commonly encountered in the literature. One should also see that it is related to the so-called inverse dielectric function $\kappa(z,z';\mathbf{k},\omega)$ by a simple relation: $\nabla^2 \kappa(z,z';\mathbf{k},\omega) + 4\pi \chi(z,z';\mathbf{k},\omega) = 0$. In general $\kappa$ takes on a much more complicated form than $\chi$. 

The response function is central to many physical processes. It has been studied mostly by means of first principles computation, in which phenomenological approximations are usually invoked~\cite{nazarov2007} and genuine surface effects are hard to be disclosed systematically. The present theory provides a physically transparent way to address these issues. 

\textbf{An identity.} The functions $B$ and $C$, and hence $P$ and $\chi$ are not independent. There is a close relation between them. We notice that the probing potential in case (i), Eq.~(\ref{2a25}), can be rewritten as $$\int^\infty_0 dq' ~ \varphi(\mathbf{K}',\omega) \cos(q'z)$$ with $$\varphi(\mathbf{K}',\omega) = (2/\pi)\left(\xi(\mathbf{k},\omega)/K^{'2}\right).$$ The induced charge density for case (i) can then be obtained as an integral over Eq.~(\ref{2a34}), i.e. $$\int^\infty_0 dq'~\chi(\mathbf{K},\mathbf{K}',\omega) \varphi(\mathbf{K}',\omega).$$ Equating this with Eq.~(\ref{2a28}), we arrive at the wanted relation, 
\begin{equation}
B(\mathbf{K},\omega) =  \frac{2}{\pi}\int^\infty_0 \frac{dq'}{K^{'2}} ~ C(\mathbf{K},\mathbf{K}',\omega), \label{bc}
\end{equation}
or equivalently, 
\begin{equation}
P(\mathbf{K},\omega) = \frac{2}{\pi}\int^\infty_0 \frac{dq'}{K^{'2}} ~ \chi(\mathbf{K},\mathbf{K}',\omega). \label{pchi}
\end{equation}
This relation shows that $\chi$ is more fundamental than $P$, namely the latter can be completely determined if the former is known while the converse is not true. 

Despite this, it is more often the function $P$ that is experimentally and theoretically analyzed, for example in energy losses of ions moving near a surface, in which cases the stimuli penetrate little or not at all into the metal so that case (i) applies. However, in experiments such as electron transmission through metal foils and where penetration is not negligible as well as optical experiments, the full structure of $\chi$ should be taken into account. To our knowledge, an analytical expression for $\chi$ has not been explicitly noted down even for the simplest model -- the DM. In the next subsection, we discuss $P$ and $\chi$ for the common models. 
 
\section{Responses within Common models}
\label{sec:2c}
The theory presented in Sec.~\ref{sec:2} is generic and applicable to any electron dynamics models, dispersive or non-dispersive. Different models lead to different expressions for $G$ and $\Omega$ as well as $B$ and $C$. In the literature, there are a few models that have been proposed and widely used for describing electron dynamics in metals. Here we discuss the most common ones, i.e. the DM, the HDM and the SRM, leaving the SCM to be systematically treated in Sec.~\ref{sec:3}. We consider the responses due to conduction electrons only. The contribution due to valence electrons is briefly discussed in Ref.~\cite{SI}.

In Table~\ref{t:1}, we summarize the defining quantities for each of the models to facilitate a quick comparison. 

\begin{table*}
\caption{\label{t:1} Summary of the defining quantities of various models within the present response theory for SIMs. DM: the classical dielectric (Drude) model. HDM: the hydrodynamic model. SRM: the specular reflection model. SCM: the semi-classical model. Denote by $\rho(\mathbf{x},t)$ the density of the charges induced in the metal by a probing electric field $\mathbf{E}_\text{probe}(\mathbf{x},t)$, and $\mathbf{E}(\mathbf{x},t)$ the electric field due to the induced charges. The current density in the metal due to $\mathbf{E}(\mathbf{x},t)$ is denoted by $\mathbf{J}(\mathbf{x},t)$. The Fourier transform of $\rho(\mathbf{x},t)$ along the surface, as defined via Eq.~(\ref{2a5}), is denoted by $\rho(z;\mathbf{k},\omega)$, where $\mathbf{k}$ is the wave vector along the surface and $\omega$ the frequency. Similar transforms are defined for other field quantities. A further cosine transform is introduced for $\rho(z;\mathbf{k},\omega)$ via Eq.~(\ref{2a10}), the $q$-th component of which is denoted by $\rho(\mathbf{K},\omega)$ with $\mathbf{K} = (\mathbf{k},q)$. The dielectric function of an infinite metal, $\epsilon$ is related to $\Omega$ by this relation: $\epsilon(K,\omega) = 1-\Omega^2(K,\omega)/\bar{\omega}^2$. The dispersion of volume plasma waves (VPWs) is given by $\epsilon(K,\omega) = 0$. Meanwhile, $G$ serves as a kernel that plays a role in $J_z(0;\mathbf{k},\omega) = \int^\infty_0 dq \left(G(\mathbf{K},\omega)/K^2\right)\rho(\mathbf{K},\omega)$. For all the models other than the SCM, $G = \omega^2_p k/\pi$, whereas for the SCM $G = \omega^2_p k/\pi + G_s$, where $G_s$ is given by Eq.~(\ref{89}). For surface plasma waves (SPWs), the most important quantity is $\epsilon_s(k,\omega) = 1 - \int^\infty_0 (dq/K^2) G(\mathbf{K},\omega)/(\Omega^2(K,\omega)-\bar{\omega}^2)$. The dispersion of SPWs is determined by $\epsilon_s(k,\omega) = 0$. The presence of $G_s$ drastically lengthens their lifetime. If the SIM is exposed to a charge of density $\rho_\text{probe}(z;\mathbf{k},\omega)$ totally residing in the vacuum, one has $\rho(\mathbf{K},\omega) = P(\mathbf{K},\omega)\xi(\mathbf{k},\omega)$, where $\xi(\mathbf{k},\omega) = 2\pi \int^0_{-\infty}dz e^{kz}\rho_\text{probe}(z;\mathbf{k},\omega)$ and $P(\mathbf{K},\omega) = \left[B(\mathbf{K},\omega)+\epsilon^{-1}_s(k,\omega)\bar{B}(\mathbf{k},\omega)\right]/(\Omega^2(K,\omega)-\bar{\omega}^2)$ with $\bar{B}(\mathbf{k},\omega) =  \int^\infty_0 (dq/K^2) B(\mathbf{K},\omega)/(\Omega^2(K,\omega)-\bar{\omega}^2)$. If the SIM is exposed to an electrostatic potential $\varphi(\mathbf{K}',\omega)\cos(q'z)$, then $\rho(\mathbf{K},\omega) = \chi(\mathbf{K},\mathbf{K}',\omega) \varphi(\mathbf{K}',\omega)$, where $\chi(\mathbf{K},\mathbf{K}',\omega) = \left[C(\mathbf{K},\mathbf{K}',\omega)+\epsilon^{-1}_s(k,\omega)\bar{C}(\mathbf{K}',\omega)\right]/(\Omega^2(K,\omega)-\bar{\omega}^2)$ is the normal density-density response function with $\bar{C}(\mathbf{K}',\omega) =  \int^\infty_0 (dq/K^2) C(\mathbf{K},\mathbf{K}',\omega)/(\Omega^2(K,\omega)-\bar{\omega}^2)$. The SRM presumes a specularly reflecting surface in the calculation of $B$ and $C$ but not in $G$, in contrast to its original contrivance. $C_s$ and $B_s$ are given by the second term of Eqs.~(\ref{94}) and (\ref{99}), respectively. The response functions $P$ and $\chi$ are not independent but related by Eq.~\ref{pchi}. They are of prime importance in many contexts but have not been analytically amenable until now. }
\begin{ruledtabular}
\begin{tabular}{c c c c c}
Quantity & DM & HDM & SRM & SCM \\ \hline
$\Omega^2(K,\omega)$ & $\omega^2_p$ & $\omega^2_p + K^2v^2_0$ & $\omega^2_p + 4\pi \bar{\omega} \mathbf{K}\cdot \mathbf{F}(\mathbf{K},\mathbf{v})/K^2$ & $\omega^2_p + 4\pi \bar{\omega} \mathbf{K}\cdot \mathbf{F}(\mathbf{K},\mathbf{v})/K^2$ \\
$G(\mathbf{K},\omega)$ & $\omega^2_p k/\pi$ & $\omega^2_p k/\pi$ & $\omega^2_p k/\pi$ & $\omega^2_p k/\pi + G_s(\mathbf{K},\omega)$ \\
$B(\mathbf{K},\omega)$ & $-\omega^2_p/4\pi$ &  $-\omega^2_p/4\pi$ &  $-\Omega^2(K,\omega)/4\pi$ &  $-\Omega^2(K,\omega)/4\pi + B_s(\mathbf{K},\omega)$ \\
$C(\mathbf{K},\mathbf{K}',\omega)$ & $-(K^2/8)\omega^2_p \delta(q-q')$ &  $-(K^2/8)\omega^2_p \delta(q-q')$ &  $-(K^2/8)\Omega^2(K,\omega) \delta(q-q')$ & $-(K^2/8)\Omega^2(K,\omega) \delta(q-q') + C_s(\mathbf{K},\mathbf{K}',\omega)$ 
\end{tabular}
\end{ruledtabular}
\end{table*}

\subsection{The local dielectric model (DM)} 
We begin the survey with the non-dispersive DM. It is the simplest model for discoursing the optical properties of metals and SPWs and often used to benchmark the validity of new methods. It is also popular for understanding electron energy loss spectroscopy and other surface related phenomena~\cite{duke1994} such as the photon drag effect~\cite{strait2019}. Here we reproduce the results known by this model but also some results which, up to our knowledge, have not been well discussed before. In the literature, the emphasis has been placed on the electromagnetic fields and the metal is viewed simply as a dielectric. Our theory deals with the charges directly. 

The DM adopts a purely local relation between the current density and the electric field, i.e. the conductivity tensor given by $\delta_{\mu\nu}\delta(z-z')\sigma_\text{DM}(\omega),$ with
\begin{equation}
\sigma_\text{DM}(\omega) = \frac{i}{\bar{\omega}}\frac{\omega^2_p}{4\pi}. \label{DM}
\end{equation} 
Here $\omega_p = \sqrt{4\pi n_0e^2/m}$ is the characteristic plasma frequency of a metal, with $n_0$ being the mean density of conduction electrons while $e$ and $m$ being the effective charge and mass of an electron, respectively. Symmetry breaking effects due to the surface are obviously excluded from this model. Thus,
\begin{equation}
\Omega = \omega_p, \quad G_s = 0, \quad G = (k/\pi)\omega^2_p. \label{2a39}
\end{equation}
The dielectric function $\epsilon(K,\omega)$ then takes on the form $$\epsilon_\text{DM}(\omega) = 1-\omega^2_p/\bar{\omega}^2,$$ which underlies the usual dielectric theory of metals. The VPW frequency is $\omega_p$ by this model. Substituting the expressions of (\ref{2a39}) into (\ref{2a19}), we find $\epsilon_s(k,\omega)$ given by
\begin{equation}
\epsilon_\text{s,DM}(\omega) = \frac{1+\epsilon_\text{DM}(\omega)}{2\epsilon_\text{DM}(\omega)},
\end{equation} 
The zero of $\epsilon_\text{s,DM}$ occurs where $$\bar{\omega} = \omega_p/\sqrt{2},$$ which is the usually quoted SPW frequency. SPWs decay in this model at a rate $\tau^{-1}$. 

\begin{figure}
\begin{center}
\includegraphics[width=0.45\textwidth]{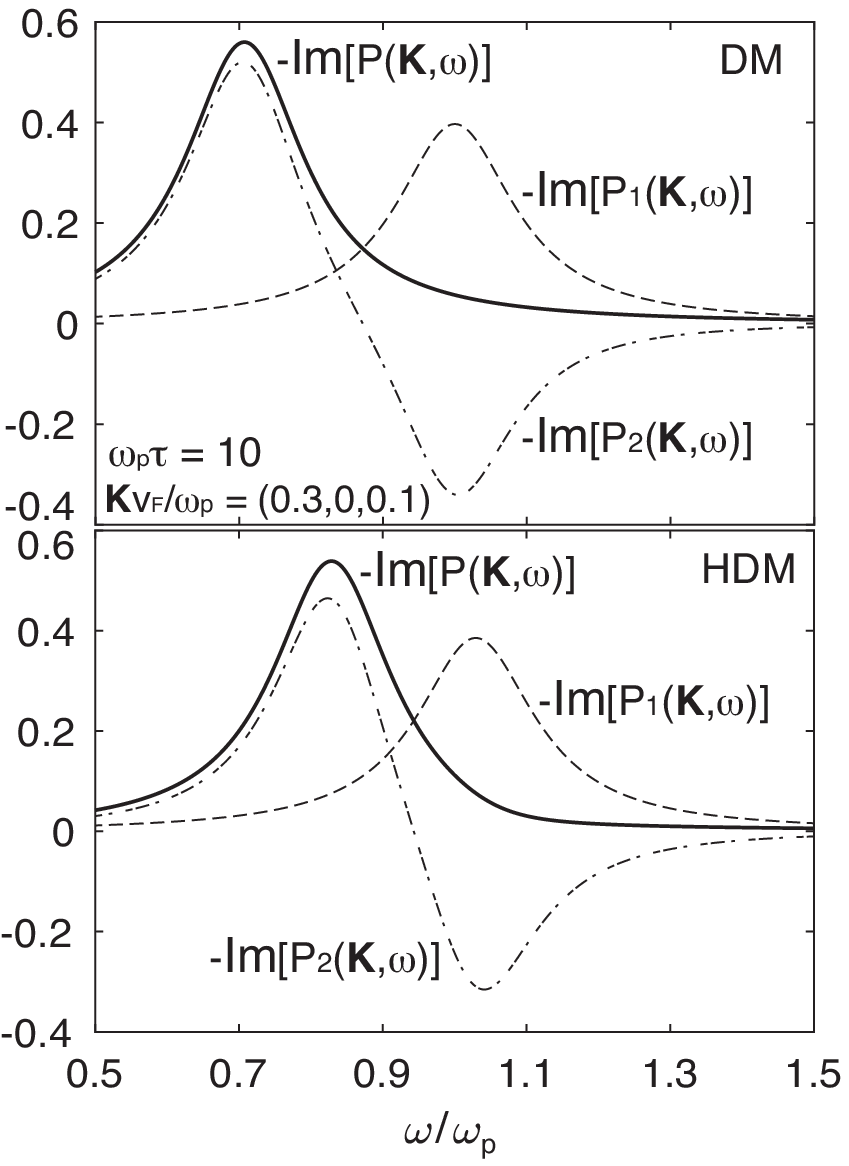}
\end{center}
\caption{The function $P(\mathbf{K},\omega) = P_1(\mathbf{K},\omega) + P_2(\mathbf{K},\omega)$ that characterizes the response to exterior charges within the DM (upper panel) and the HDM (lower panel). There is nearly complete cancellation between $P_1$ and $P_2$ near the VPW resonances and only the SPW peak appears in $P$. Parameters are the same in both panels. \label{fig:P}}
\end{figure}

Let us examine the responses to exterior charges as described in Sec.~\ref{sec:2b}. In the first place, we have $$\nabla\cdot \mathbf{J}_\text{probe}(z;\mathbf{k},\omega) = 4\pi \sigma_\text{DM}(\omega) \rho_\text{ext}(z;\mathbf{k},\omega),$$ which vanishes in the metal by definition. It follows that
\begin{equation}
S_\text{probe}(\mathbf{K},\omega) = i\bar{\omega}J_\text{probe,z}(0;\mathbf{k},\omega) = i\bar{\omega}\sigma_\text{DM}(\omega) \xi(\mathbf{k},\omega), \label{2a42}
\end{equation}
where we have used Eq.~(\ref{2a26}). This leads to
\begin{equation}
B(\mathbf{K},\omega) = i\bar{\omega}\sigma_\text{DM}(\omega) = -\frac{\omega^2_p}{4\pi}, \quad P_1(\mathbf{K},\omega) = \frac{1-\epsilon_\text{DM}}{4\pi \epsilon_\text{DM}}. \label{2a43}
\end{equation}
Similarly, we find
\begin{equation}
\bar{B}(\mathbf{k},\omega) = \frac{\omega^2_p}{4\pi}\frac{\omega^2_p}{2\bar{\omega}^2}\frac{1}{\epsilon_\text{DM}(\omega)}, \quad P_2(\mathbf{K},\omega) = P_1(\mathbf{K},\omega) ~ \frac{\epsilon_\text{DM}-1}{\epsilon_\text{DM} + 1}. 
\end{equation}
Combining $P_1$ and $P_2$, we arrive at
\begin{equation}
P(\omega):= P(\mathbf{K},\omega) = P_1(\mathbf{K},\omega)/\epsilon_\text{s,DM}(\omega) = \frac{1}{2\pi} \frac{1-\epsilon_\text{DM}(\omega)}{1+\epsilon_\text{DM}(\omega)}, \label{2a44}
\end{equation}
We see that, although $P_1$ features a resonance near the zero of $\epsilon(\omega)$, $P$ does not display such a resonance. Instead, only the resonance near the zero of $\epsilon_s$ exists with $P$. As aforementioned, this is due to the cancellation between $P_1$ and $P_2$ near the VPW frequency, as is displayed in the upper panel of Fig. \ref{fig:P} for an illustration. It is clear that ABCs are incompatible with this model.

Equation (\ref{2a44}) is one of the most used results for analyzing surface excitations and other surface phenomena such as the energy absorption of grazing particles and photon drag effect. 

As for the induced charge density in this case, we see that $\rho(\mathbf{K},\omega)$ does not depend on $q$ in this model, leading to $\rho(z;\mathbf{k},\omega) = \rho_s \Theta'(z)$ purely localized on the surface, where $\rho_s = P(\omega)\xi(\mathbf{k},\omega)$ is the areal surface charge density. 

The responses to an electrostatic potential -- case (ii) -- can be similarly dealt with. By Eq.~(\ref{2a32}), we deduce that $$J_\text{probe,z}(0;\mathbf{k},\omega) = \sigma_\text{DM}(\omega)E_\text{probe,z}(0;\mathbf{k},\omega) = 0.$$ In addition, $$\nabla\cdot \mathbf{J}_\text{probe}(z;\mathbf{k},\omega) = \sigma_\text{DM}(\omega) \varphi(\mathbf{K}',\omega)K^{'2}\cos(q'z).$$ The corresponding $S_\text{probe}$ is obtained as 
\begin{equation}
S_\text{probe}(\mathbf{K},\omega) = -\varphi(\mathbf{K}',\omega)\left(K^{'2}/8\right)\omega^2_p \delta(q-q'),
\end{equation}
which leads to
\begin{equation}
C(\mathbf{K},\mathbf{K}',\omega) = - \left(K^{2}/8\right)\omega^2_p \delta(q-q'). \label{2a47}
\end{equation}
Substituting this in Eq.~(\ref{2a31}), we arrive at
\begin{equation}
\bar{C}(\mathbf{K}',\omega) = \frac{k\omega^2_p}{8\pi}\frac{\omega^2_p}{\bar{\omega}^2}\frac{1}{\epsilon_\text{DM}(\omega)}. 
\end{equation}
Finally, 
\begin{equation}
\chi_1(\mathbf{K},\mathbf{K}',\omega) = \frac{K^2}{8}\frac{1-\epsilon_\text{DM}}{\epsilon_\text{DM}} \delta(q-q')
\end{equation}
and
\begin{equation}
\chi_2(\mathbf{K},\mathbf{K}',\omega) = \frac{k}{4\pi}\frac{\epsilon_\text{DM}-1}{\epsilon_\text{DM}+1}\frac{1-\epsilon_\text{DM}}{\epsilon_\text{DM}}. 
\end{equation}
Combined, they produce
\begin{equation}
\chi(\mathbf{K},\mathbf{K}',\omega) = \frac{1-\epsilon_\text{DM}}{\epsilon_\text{DM}}\left(\frac{K^2}{8}\delta(q-q') - \frac{k}{2}~P(\omega)\right), \label{chiDM}
\end{equation}
This result is not widely known, though an equivalent but much more involved expression has been written down in Ref.~\cite{aminov2001} for the non-local dielectric function. Most authors have considered only the responses due to SPWs, i.e. the second term in Eq.~(\ref{chiDM}). 

Unlike $P$, $\chi$ contains resonances of both VPWs and SPWs. Obviously, $\chi$ and $P$ fulfill the relation (\ref{pchi}). 

\subsection{The hydrodynamic model (HDM)}
The DM assumes a local dependence of the current density on the electric field. In recent years there has seen lots of interest in the HDM, which is a slight extension of the DM by inclusion of some non-local effects. There are several paths, which are not always equivalent, to the HDM~\cite{noteHDM}. Here we use the fluid mechanics approach, by which the current density is given by
\begin{equation}
\mathbf{J}(z;\mathbf{k},\omega) = \frac{i}{\bar{\omega}}\left(\frac{\omega^2_p}{4\pi}\mathbf{E}(z;\mathbf{k},\omega) - v^2_0\nabla \rho(z;\mathbf{k},\omega)\right), \label{2a52}
\end{equation}
where $v_0$ is a parameter. The first term here is the same as in the DM, while the second one due to inter-electron interactions gives rise to non-local responses. In addition,
\begin{equation}
\mathbf{J}_\text{probe}(z;\mathbf{k},\omega) = \frac{i}{\bar{\omega}}\frac{\omega^2_p}{4\pi} \mathbf{E}_\text{probe}(z;\mathbf{k},\omega), \label{2a53}
\end{equation}
which has the same form as in the DM. With these two relations, one can show that
\begin{equation}
\Omega^2_\text{HDM}(K) = \omega^2_p + v^2_0K^2, \quad G_s = 0, \quad G = (k/\pi) \omega^2_p. 
\end{equation}
The dielectric function is then given by~\cite{noteHDM}
\begin{equation}
\epsilon_\text{HDM}(K,\omega) = 1-\frac{\Omega^2_\text{HDM}(K)}{\bar{\omega}^2}. \label{epHDM}
\end{equation}
The VPW dispersion is given by $\Omega_\text{HDM}(K)$. The corresponding $\epsilon_s(k,\omega)$ is found to be
\begin{equation}
\epsilon_{s,\text{HDM}}(k,\omega) = 1 + \frac{\omega^2_p}{2\bar{\omega}^2}\frac{k}{\pi}\int^\infty_{-\infty} \frac{dq}{K^2} \frac{1}{\epsilon_\text{HDM}(K,\omega)}, \label{eshdm}
\end{equation}
whose zeros give the SPW dispersion in the HDM. 

Equation (\ref{eshdm}) recovers $\epsilon_\text{s,DM}$ in the limit $v_0 = 0$. By solving the equation $\epsilon_\text{s,HDM} = 0$ we find that the SPW dispersion relation in the HDM, approximately given by $$\left(\omega_p/\sqrt{2}\right)\left(1+\alpha kv_0/\omega_p\right)$$ exhibits a linear $k$ dependence. Here $\alpha$ is a constant of the order of unity. As thoroughly discussed in Ref.~\cite{deng2018}, the widely adopted treatment of SPWs within the HDM is incorrect and the DM cannot be recovered in that treatment.

The responses to exterior charges can easily be obtained using Eq.~(\ref{2a53}). Obviously $S_\text{probe}$ and $B(\mathbf{K},\omega)$ are the same as in the DM, see Eqs.~(\ref{2a42}) and (\ref{2a43}), while
\begin{equation}
\bar{B}(k,\omega) = \frac{\omega^2_p}{2\bar{\omega}^2}\frac{\omega^2_p}{4\pi}\frac{k}{\pi}\int^\infty_{-\infty} \frac{dq}{K^2}\frac{1}{\epsilon_\text{HDM}(K,\omega)} = \frac{\omega^2_p}{4\pi}\left(\epsilon_\text{s,HDM} - 1\right). \label{2a57}
\end{equation} 
In obtaining the second equality we have used Eq.~(\ref{eshdm}). We thus find
\begin{equation}
P_1(\mathbf{K},\omega) = \frac{\omega^2_p}{\bar{\omega}^2}\frac{1}{4\pi \epsilon_\text{HDM}(K,\omega)},
\end{equation}
and 
\begin{equation}
P_2(\mathbf{K},\omega) = \frac{\omega^2_p}{4\pi \bar{\omega}^2}\frac{1-\epsilon_\text{s,HDM}(k,\omega)}{\epsilon_\text{HDM}(K,\omega)\epsilon_\text{s,HDM}(k,\omega)}. 
\end{equation}
Combined, they yield
\begin{equation}
P(\mathbf{K},\omega) = P_1(\mathbf{K},\omega)/\epsilon_\text{s,HDM}(k,\omega), 
\end{equation}
which reduces in the limit $v_0 = 0$ to that for the DM. Again there is nearly perfect cancellation between $P_1$ and $P_2$ near the VPW resonances, as seen in the lower panel of Fig.~\ref{fig:P}. The induced charge density $\rho(\mathbf{K},\omega)$ now depends on $K$ via $\epsilon^{-1}_\text{HDM}(K,\omega)$. For $\omega < \omega_p$, the charges are localized within a layer of thickness around $v_0/\omega_p$. 

As for the responses to an electrostatic potential, we see that $C(\mathbf{K},\mathbf{K}',\omega)$ is also the same as in the DM, given by Eq.~(\ref{2a47}). It follows that 
\begin{equation}
\bar{C}(\mathbf{K}',\omega) = \frac{k\omega^2_p}{8\pi}\frac{\omega^2_p}{\bar{\omega}^2}\frac{1}{\epsilon_\text{HDM}(K',\omega)}. 
\end{equation}
Combining these expressions yields
\begin{equation}
\chi_1(\mathbf{K},\mathbf{K}',\omega) = \frac{K^2}{8}\frac{\omega^2_p}{\bar{\omega}^2}\frac{1}{\epsilon_\text{HDM}(K,\omega)}\delta(q-q') 
\end{equation}
and
\begin{equation}
\chi_2(\mathbf{K},\mathbf{K}',\omega) = -\frac{\omega^2_p/\bar{\omega}^2}{\epsilon_\text{HDM}(K,\omega)}\frac{\omega^2_p/\bar{\omega}^2}{\epsilon_\text{HDM}(K',\omega)}\frac{k/8\pi}{\epsilon_\text{s,HDM}(k,\omega)}.
\end{equation}
Combined, they lead to
\begin{eqnarray}
&~& \chi(\mathbf{K},\mathbf{K}',\omega) = \frac{1}{8}\frac{\omega^2_p/\bar{\omega}^2}{\epsilon_\text{HDM}(K,\omega)} \nonumber \\ &~& \quad \quad \times \left(K^2\delta(q-q')-\frac{\omega^2_p/\bar{\omega}^2}{\epsilon_\text{HDM}(K',\omega)}\frac{k/\pi}{\epsilon_\text{s,HDM}(k,\omega)}\right).
\end{eqnarray}
Up to our knowledge, these functions have never been discussed in the literature, even though the HDM is a popular model for electron dynamics~\cite{barton1979}. 

\subsection{The specular reflection model (SRM)}
In the HDM, $\Omega$ is approximated by $\Omega_\text{HDM}$, which is valid only for small $K$. The next natural step is to use the exact form of $\Omega$ so that the dielectric function $\epsilon(K,\omega)$ becomes exact, while still neglecting the symmetry breaking effects, i.e. one approximates 
\begin{equation}
G_s = 0, \quad G = (k/\pi)\omega^2_p. 
\end{equation}
The ensuing $\epsilon_s(k,\omega)$ then takes on the following form
\begin{equation}
\epsilon_{s,\text{SRM}}(k,\omega) = 1 + \frac{\omega^2_p}{2\bar{\omega}^2}\frac{k}{\pi}\int^\infty_{-\infty} \frac{dq}{K^2} \frac{1}{\epsilon(K,\omega)}. 
\end{equation}
The VPW dispersion relation is obtained by solving the equation that $\bar{\omega} = \Omega(K,\omega)$ while the SPW dispersion relation by the following equation
\begin{equation}
\epsilon_\text{s,SRM}(k,\omega) = 0. \label{2a67}
\end{equation}
which is nothing but the SRM equation for SPWs first proposed by Ritchie and Marusak~\cite{ritchie1966} in 1966. The present derivation makes it clear that the SRM can be regarded as an extension of the HDM. In contrast to its original contrivance, the SRM does not simply assume a specularly reflecting surface in actuality; otherwise, one would have no surface contribution and Eq.~(\ref{2a67}) would not have been reached. More discussions on the logical structure of this widely used SRM are given in Ref.~\cite{SI}. As with the DM and the HDM, the SRM also excludes symmetry breaking effects from $G$. 

The responses within the SRM will be briefly discussed in the next section, in parallel with the SCM. The quantities $B$ and $C$ are quoted here. They are given by 
\begin{eqnarray}
&~& C(\mathbf{K},\mathbf{K}',\omega) = -K^2\Omega^2(K,\omega)\delta(q-q')/8, \label{srmchi} \\ &~& B(\mathbf{K},\omega) = - ~ \Omega^2(K,\omega)/4\pi, \label{srmb}
\end{eqnarray}
which are direct generalizations of the DM and HDM counterparts. Now 
\begin{equation}
\bar{B}(k,\omega) = \frac{1}{4\pi}\frac{\omega^2_p}{2\bar{\omega}^2}\frac{k}{\pi}\int^\infty_{-\infty} \frac{dq}{K^2}\frac{\Omega^2(K,\omega)}{\epsilon(K,\omega)}, 
\end{equation}
which may be rewritten as $$\bar{B}(k,\omega) = \left(\overline{\Omega^2}/4\pi\right)\left(\epsilon_\text{s,SRM}(k,\omega)-1\right),$$ where $\overline{\Omega^2}$ is defined by 
\begin{equation}
\overline{\Omega^2} = \int^\infty_{-\infty} \frac{dq}{K^2}\frac{\Omega^2(K,\omega)}{\epsilon(K,\omega)}/\int^\infty_{-\infty} \frac{dq}{K^2}\frac{1}{\epsilon(K,\omega)},\label{barWS}
\end{equation}
which is plotted in Fig.~\ref{fig:WS} (b). From these we obtain
\begin{eqnarray}
P_1(\mathbf{K},\omega) &=&\frac{1-\epsilon(K,\omega)}{4\pi\epsilon(K,\omega)} = \frac{\Omega^2(K,\omega)}{\bar{\omega}^2}\frac{1}{4\pi\epsilon(K,\omega)}, \\
P_2(\mathbf{K},\omega) &=& \frac{\overline{\Omega^2}}{\bar{\omega}^2}\frac{1}{4\pi\epsilon(K,\omega)}\frac{1-\epsilon_\text{s,SRM}(k,\omega)}{\epsilon_\text{s,SRM}(k,\omega)},
\end{eqnarray}
which closely resemble those in the HDM. If we approximate $\Omega^2 \approx \overline{\Omega^2}$, this leads to
\begin{equation}
P(\mathbf{K},\omega) \approx P_1(\mathbf{K},\omega)/\epsilon_\text{s,SRM}(k,\omega). 
\end{equation}
This may be a good approximation for small $Kv_F/\bar{\omega}$, where $\Omega^2$ shows little dispersion as discussed in the next section. 

\section{Responses by the semi-classical model}
\label{sec:3}
In the SCM one calculates the electrical responses due to conduction electrons in terms of a distribution function $f(\mathbf{x},\mathbf{v},t)$ defined in the single-particle phase space. Here $\mathbf{v} = (\mathbf{v}_\parallel,v_z)$ denotes the velocity of electrons, where $\mathbf{v}_\parallel = (v_x,v_y)$ is the planar component. As usual, we write the function as a sum of an equilibrium part $f_0(\varepsilon(\mathbf{v}))$ and a non-equilibrium part $g(\mathbf{x},\mathbf{v},t)$. $f_0(\varepsilon)$ is taken to be the Fermi-Dirac function at zero temperature. $\varepsilon(\mathbf{v}) = mv^2/2$ is the energy dispersion of the conduction band. Within the relaxation time approximation and the regime of linear responses, the Fourier components of $g(\mathbf{x},\mathbf{v},t)$ satisfy the following Boltzmann's equation
\begin{equation}
    \left(\lambda^{-1} + \partial_z\right)g(\mathbf{v},z;\mathbf{k},\omega) + ef'_0(\varepsilon)\mathbf{v}\cdot\mathbf{E}(z;\mathbf{k},\omega)/v_z = 0. \label{3.1}
\end{equation}
Here $\lambda = iv_z/\tilde{\omega}$ with $\tilde{\omega} = \bar{\omega} - \mathbf{k}\cdot \mathbf{v}_\parallel$ and $f'_0 = \partial_{\varepsilon}f_0(\varepsilon)$. The electric field $\mathbf{E}(z;\mathbf{k},\omega)$ is not specified here: it can be due to the induced charges or the probing field or the total field. As dictated by causality~\cite{deng2017a}, $\gamma_0 = $~Im$(\bar{\omega})$ must be non-negative and the general solution is then given by 
\begin{equation}
g(\mathbf{v},z;\mathbf{k},\omega) = e^{-\frac{z}{\lambda}}\left(C_{\mathbf{k}\omega}(\mathbf{v})-\frac{ef'_0\mathbf{v}}{v_z}\cdot \int^z_0~dz'~e^{\frac{z'}{\lambda}}~\mathbf{E}(z';\mathbf{k},\omega)\right), \label{3.2}
\end{equation}
where $C_{\mathbf{k}\omega}(\mathbf{v}) = g(\mathbf{v},0;\mathbf{k},\omega)$ is the non-equilibrium deviation on the surface to be determined by boundary conditions. We require $g(\mathbf{v},z;\mathbf{k},\omega)=0$ distant from the surface, i.e. $z\rightarrow\infty$. For electrons moving away from the surface, $v_z>0$, this condition is automatically fulfilled. For electrons moving toward the surface, $v_z<0$, it leads to
\begin{equation}
C_{\mathbf{k}\omega}(\mathbf{v}) =\frac{ef'_0\mathbf{v}}{v_z}\cdot \int^{\infty}_0~dz'~e^{z'/\lambda}\mathbf{E}(z';\mathbf{k},\omega), \quad v_z<0, 
\end{equation}
yielding
\begin{equation}
g(\mathbf{v},z;\mathbf{k},\omega) = \frac{ef'_0\mathbf{v}}{v_z}\cdot \int^{\infty}_zdz' ~e^{\frac{z'-z}{\lambda}}~\mathbf{E}(z';\mathbf{k},\omega), \quad v_z<0. 
\end{equation}
To determine $C_{\mathbf{k}\omega}(\mathbf{v})$ for $v_z>0$, the boundary condition at $z=0$ has to be used, which, whoever, depends on the surface scattering properties. We adopt a simple picture that was first conceived by Fuchs~\cite{fuchs1938} and afterwards widely used in the study of for instance anomalous skin effect~\cite{reuter1948,ziman,abrikosov}. According to this picture a fraction $p$ -- the \textit{Fuchs} parameter varying between zero and unity -- of the electrons impinging on the surface are specularly reflected back, i.e. 
\begin{equation}
g(\mathbf{v},z=0;\mathbf{k},\omega)=p~g(\mathbf{v}_-,z=0;\mathbf{k},\omega),  \label{3.5} 
\end{equation}
where $\mathbf{v}_- = (v_x,v_y,-v_z)$ with $v_z\geq 0$. It follows that 
\begin{equation}
C_{\mathbf{k}\omega}(\mathbf{v}) = - p~\frac{ef'_0\mathbf{v}_-}{v_z}\cdot \int^{\infty}_0dz' ~e^{-\frac{z'}{\lambda}}~\mathbf{E}(z';\mathbf{k},\omega), \quad v_z\geq0. \label{3.6}
\end{equation} 
Equations (\ref{3.2}) - (\ref{3.6}) fully specify the distribution function for the electrons due to a field. 

The corresponding current density is calculated in the usual way, 
\begin{equation}
\mathbf{J}(z;\mathbf{k},\omega) = \left(\frac{m}{2\pi\hbar}\right)^3\int d^3\mathbf{v} ~e\mathbf{v}~g(\mathbf{v},z;\mathbf{k},\omega). \label{3.7} 
\end{equation} 
Surface roughness enters the responses through the reflected electrons of fraction $p$. It is guaranteed that $J_z(0;\mathbf{k},\omega) = 0$ for specularly reflecting surfaces ($p=1$). Nevertheless, the charge density is not given by $$\tilde{\rho}(\mathbf{x},t)=(m/2\pi\hbar)^3~e^{i(kx-\omega t)}\int d^3\mathbf{v}~eg(\mathbf{v},z).$$ The reason is because Eq.~(\ref{3.1}) and hence the as-obtained $g(\mathbf{v},z)$ is for the bulk region and not valid on the surface~\cite{deng2020b}, since it involves no surface potentials, as explained in Sec.~\ref{sec:2} and in previous work~\cite{deng2017c}. Actually, $\mathbf{J}(\mathbf{x},t)$ and $\tilde{\rho}(\mathbf{x},t)$ obey the equation $$(\partial_t+1/\tau)\tilde{\rho}(\mathbf{x},t)+\partial_{\mathbf{x}}\cdot\mathbf{J}(\mathbf{x},t)=0$$ rather than the equation of continuity [c.f. Eq.~(\ref{2a2})], thus automatically but incorrectly embodying the condition that $J_z(0)=0$. This underlies the incorrect conclusion drawn by Harris~\cite{harris1979} and calls into question many other works such as Ref.~\cite{principi2018}. That Eq.~(\ref{3.1}) is for the bulk also justifies $f_0$ being simply the Fermi-Dirac function, since $f_0$ is the bulk equilibrium distribution without the impact of surface potential. We should also remark that Eq.~(\ref{3.1}) assumes a global relaxation term. More accurately, it may be replaced with a local relaxation term. However, the difference is a higher-order effect~\cite{fuchs1968}, which is negligible in the electrostatic limit concerned in the present work. 

\subsection{Expressions for $\Omega(K,\omega)$ and $G(\mathbf{K},\omega)$}
\label{sec:3a}
Now we specify to the case where the field in the distribution function is due to the induced charges. We substitute the expressions of $\mathbf{E}(z;\mathbf{k},\omega)$, i.e. Eqs.~(\ref{ex}) and (\ref{ez})  into (\ref{3.2}) - (\ref{3.6}) and perform the integration over $z'$. The resulting distribution function $g(\mathbf{v},z;\mathbf{k},\omega)$ may be split in two parts, one denoted by $g_b(\mathbf{v},z)$ and the other by $g_s(\mathbf{v},z)$. They are given by
\begin{eqnarray}
&~&g_b(\mathbf{v},z;\mathbf{k},\omega) = -ef'_0\int^\infty_0 dq ~ \frac{4\rho_q}{K^2}\times \label{gb} \\ &~& \left[F_+(\mathbf{K},\bar{\omega},\mathbf{v})\cos(qz)+iF_-(\mathbf{K},\bar{\omega},\mathbf{v})\sin(qz)-F_0(\mathbf{k},\bar{\omega},\mathbf{v})e^{-kz}\right], \nonumber
\end{eqnarray}
where we have introduced the following functions,
\begin{equation}
F_\pm(\mathbf{K},\bar{\omega},\mathbf{v}) = \frac{\mathbf{K}\cdot\mathbf{v}}{\bar{\omega} - \mathbf{K}\cdot\mathbf{v}}\pm \frac{\mathbf{K}\cdot\mathbf{v}_-}{\bar{\omega} - \mathbf{K}\cdot\mathbf{v}_-}.
\end{equation}
$F_\pm$ is an even/odd function of $v_z$. They signify the bulk responses in the presence of two counter-propagating waves $e^{\pm iqz}$ superposed in/out of phase with equal weights. In addition, 
\begin{equation}
F_0(\mathbf{k},\bar{\omega},\mathbf{v}) = \frac{\mathbf{k}^*\cdot\mathbf{v}}{\bar{\omega} - \mathbf{k}^*\cdot\mathbf{v}} = \sum^\infty_{l=1}\left(\frac{\mathbf{k}^*\cdot\mathbf{v}}{\bar{\omega}}\right)^l, \quad \mathbf{k}^* = (\mathbf{k},ik), \end{equation}
which stems from the exponential term of the electric field. The other part is given by
\begin{eqnarray}
&~&g_s(\mathbf{v},z;\mathbf{k},\omega) = \Theta(v_z)(-ef'_0)e^{i\frac{\bar{\omega}z}{v_z}}\int^\infty_0 dq\frac{4\rho_q}{K^2}\times \label{gs} \\ &~& \quad \left[ F_0(\mathbf{k},\bar{\omega},\mathbf{v})-p F_0(\mathbf{k},\bar{\omega},\mathbf{v}_-) + (p-1)F_+(\mathbf{K},\bar{\omega},\mathbf{v})\right]. \nonumber
\end{eqnarray}

One may also obtain $g_b$ by the arguments of Ritchie and Marusak leading to the SRM~\cite{ritchie1966} or directly by solving Boltzmann's equation for an infinite system. This part gives exactly the responses for an infinite system. It is independent of surface properties, i.e. showing no dependence on the \textit{Fuchs} parameter $p$, and the electrons incident on the surface (i.e. with $v_z<0$) and those departing it (i.e. with $v_z>0$) appear on equal footing in its expression. If we keep only $g_b$, the SRM equation (\ref{2a67}) will be revisited, making it evident that the SRM does not correspond to the limit of $p=1$ (specularly reflecting surface). Instead, it corresponds to the neglect of $g_s$. In this sense, 'SRM' is a misnomer for the model. 
  
On the contrary, $g_s$ signifies pure symmetry breaking effects: it exists only for departing electrons, as indicated by the Heaviside function $\Theta(v_z)$ in its expression, and it depends on $p$ and thus reflects on surface scattering properties. Another important feature of $g_s$ lies in its simple dependence on $z$, i.e. $g_s \propto e^{i\tilde{\omega}z/v_z}$. As we reasoned in Refs.~\cite{deng2018,deng2017a,deng2017b,deng2017c}, this factor in accord with causality implies $\gamma_0\geq0$ and an intrinsic instability of the metal against SPWs only to be stabilized by thermal electronic collisions. 

Now we can easily find the current density and the expressions of $\Omega$ and $G$. Let us split the current density in two parts, $\mathbf{J}(z;\mathbf{k},\omega) = \mathbf{J}_b(z;\mathbf{k},\omega) + \mathbf{J}_s(z;\mathbf{k},\omega)$, where $\mathbf{J}_{b/s}(z;\mathbf{k},\omega)$ are defined via Eq.~(\ref{3.7}) with $g(\mathbf{v},z;\mathbf{k},\omega)$ replaced by $g_{b/s}(\mathbf{v},z;\mathbf{k},\omega)$. For small $kv_F/\bar{\omega}$, we may retain only the first term in the series of $F_0(\mathbf{k},\bar{\omega},\mathbf{v})$; Actually the next order contribution comes from the third term rather than the second and therefore negligible. We find that 
\begin{equation}
    \mathbf{J}_b(z;\mathbf{k},\omega) = \sigma_\text{DM}(\omega) \mathbf{E}(z;\mathbf{k},\omega) + \mathbf{J}_\text{SRM}(z;\mathbf{k},\omega).
\end{equation}
Here $\mathbf{J}_\text{SRM}(z;\mathbf{k},\omega)$ is responsible for the extension made in the SRM beyond the DM. It is given by
\begin{eqnarray}
J_{\text{SRM},x/y}(z;\mathbf{k},\omega) &=& \int \mathcal{D}q\mathcal{D}^3\mathbf{v}~v_{x/y} F'_+(\mathbf{K},\bar{\omega},\mathbf{v})\cos(qz), \\
J_{\text{SRM},z}(z;\mathbf{k},\omega) &=& i ~ \int \mathcal{D}q\mathcal{D}^3\mathbf{v}~v_z F'_-(\mathbf{K},\bar{\omega},\mathbf{v})\sin(qz),
\end{eqnarray}
where we have defined a short-hand
$$\int \mathcal{D}q\mathcal{D}^3\mathbf{v} ... = \left(\frac{m}{2\pi\hbar}\right)^3 \int^\infty_0 dq~\frac{4\rho_q}{K^2} \int d^3\mathbf{v}\left(-e^2f'_0\right) ...$$
together with these functions $$F'_\pm(\mathbf{K},\bar{\omega},\mathbf{v}) = \frac{1}{2}\left[\frac{(\mathbf{K}\cdot\mathbf{v})^2}{1 - \mathbf{K}\cdot\mathbf{v}/\bar{\omega}}\pm \frac{(\mathbf{K}\cdot\mathbf{v}_-)^2}{1 - \mathbf{K}\cdot\mathbf{v}_-/\bar{\omega}}\right].$$ See that $J_{\text{SRM},z}(0;\mathbf{k},\omega)\equiv 0$, which means that $\mathbf{J}_\text{SRM}$ makes no contribution to $G$. One thus concludes that $$G_b = (k/\pi)\omega^2_p$$ as with the DM and other models. 

By their definitions, Eqs.~(\ref{2a8}), (\ref{2a13}) and (\ref{2a14}), we directly find that 
\begin{equation}
    \Omega^2(K,\omega) = \omega^2_p + \frac{4\pi\bar{\omega}\mathbf{K}\cdot\mathbf{F}(\mathbf{K},\bar{\omega})}{K^2}, \label{3.15}
\end{equation}
where $\mathbf{F}(\mathbf{K},\bar{\omega})$ is an odd function of $\bar{\omega}$ and given by
\begin{equation}
    \mathbf{F}(\mathbf{K},\bar{\omega}) = \left(\frac{m}{2\pi\hbar}\right)^3 \int d^3\mathbf{v}\left(-e^2f'_0\right)\left(\frac{\mathbf{K}\cdot\mathbf{v}}{\bar{\omega}}\right)^2\frac{\mathbf{v}}{1-\mathbf{K}\cdot\mathbf{v}/\bar{\omega}}. 
\end{equation}
See that $\mathbf{K}\cdot\mathbf{F}$ does not depend on the direction of $\mathbf{K}$. Additionally, we have 
\begin{eqnarray}
    &~& G_s(\mathbf{K},\omega) = 4i\bar{\omega}\left(\frac{m}{2\pi\hbar}\right)^3\int_>d^3\mathbf{v}~v_z\left(-e^2f'_0\right)\times \label{Gsa} \\
    &~& \quad \quad \quad \left[ F_0(\mathbf{k},\bar{\omega},\mathbf{v})-p F_0(\mathbf{k},\bar{\omega},\mathbf{v}_-) + (p-1)F_+(\mathbf{K},\bar{\omega},\mathbf{v})\right], \nonumber
\end{eqnarray}
which strongly depends on $p$. Here the integral is restricted to $v_z\geq 0$, as indicated by the symbol '$>$'. 

\begin{figure*}
\begin{center}
\includegraphics[width=0.9\textwidth]{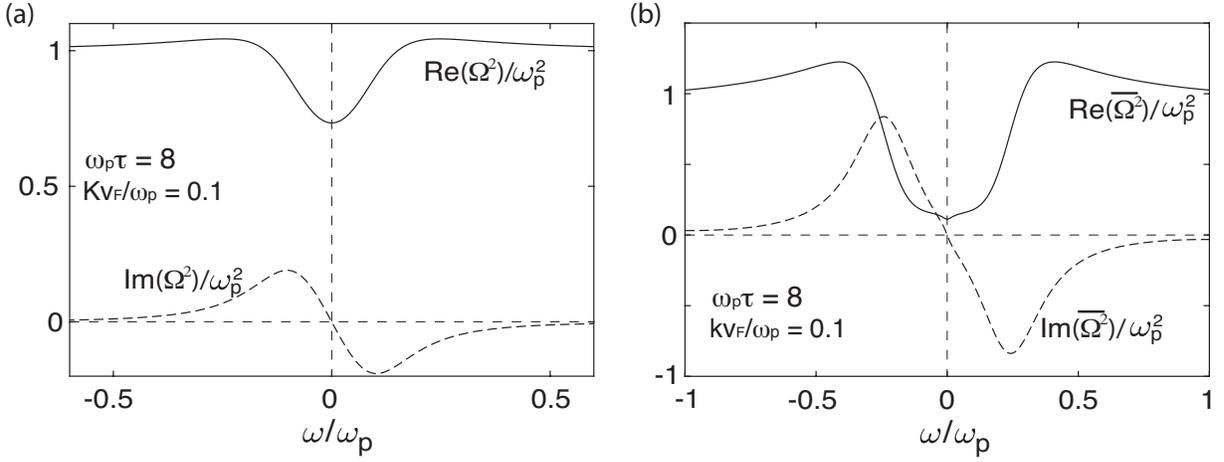}
\end{center}
\caption{Frequency dependence of (a) $\Omega^2(K,\omega)$ [Eq.~(\ref{3.85})] and (b) $\overline{\Omega^2}(k,\omega)$ [Eq.~(\ref{barWS})]. Re$[\Omega^2(K,\omega)]$ is even in $\omega$ whereas Im$[\Omega^2(K,\omega)]$ is odd in $\omega$. At $\omega = 0$, $\Omega$ is real. $\Omega^2$ depends on $K$ and $\bar{\omega}$ via the combination $Kv_F/\bar{\omega}$, rather than individually. Similar properties hold for $\overline{\Omega^2}(k,\omega)$. For large $\omega$, these two quantities become comparable. \label{fig:WS}}
\end{figure*}

The second term in Eq.~(\ref{3.15}) is generally complex even in the collision\textit{less} limit where $\tau^{-1}$ is vanishingly small, due to a pole at $\bar{\omega} = \mathbf{K}\cdot\mathbf{v}$ in the integrand in $\mathbf{F}$. The imaginary part of $\Omega^2$ gives rise to Landau damping, i.e. the damping due to the excitation of particle-hole pairs. Its real part approximates $$\omega^2_p + \frac{3}{5} K^2v^2_F$$ for small $K$, which revisits $\Omega_\text{HDM}$ with $v_0 = \sqrt{\frac{3}{5}} ~ v_F$. The integral in the expression of $\mathbf{F}$ can be partially performed. Doing this leads to
\begin{equation}
\Omega^2(K,\omega) = \omega^2_p \left(1 + \frac{3}{2}\frac{Kv_F}{\bar{\omega}} \int^1_{-1} dr ~ \frac{r^3}{1-rKv_F/\bar{\omega}}\right). \label{3.85}
\end{equation}
It shows that $\Omega$ depends on $K$ and $\omega$ not individually, but only through the ratio $Kv_F/\bar{\omega}$. In Fig.~\ref{fig:WS} (a), $\Omega$ is plotted, where it is seen that the real (imaginary) part of $\Omega^2$ is even (odd) in $\omega$, a property that can be rigorously proved by use of the relation that $\mathbf{F}(\mathbf{K},\bar{\omega}) + \mathbf{F}(\mathbf{K},-\bar{\omega}) = 0$. The imaginary part displays a minimum on the physical (positive) frequency side, due to particle-hole excitations produced at $\omega = Kv_F$ that is responsible for Landau damping. 

\begin{figure}
\begin{center}
\includegraphics[width=0.45\textwidth]{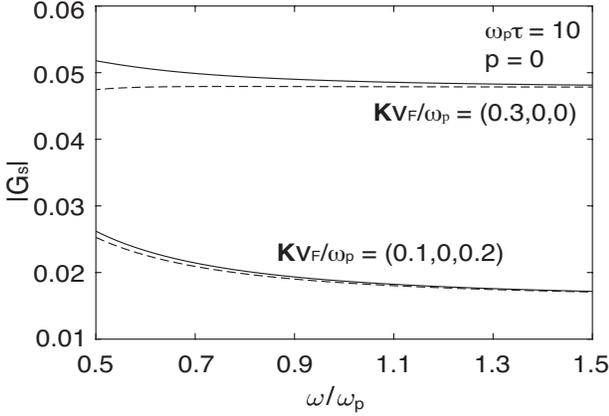}
\end{center}
\caption{Illustration of $G_s(\mathbf{K},\omega)$, which contains symmetry breaking effects and disappears from all the models except the SCM. Solid line: Eq.~(\ref{Gsa}). Dashed line: Eq.~(\ref{89}). \label{fig:G}}
\end{figure}

A crucial improvement of the SCM over the SRM comes through the quantity $G_s(\mathbf{K},\omega)$. In the SRM and its descendents, $G_s = 0$ and no symmetry breaking effects are present. As shown in Refs.~\cite{deng2018,deng2017a,deng2017b,deng2017c}, thanks to $G_s$, an instability of the metal might be induced at some critical point, where SPWs become lossless with infinitely long lifetime -- a highly desirable attribute in plasmonics and other practical areas of SPWs. For small $kv_F/\bar{\omega}$, we may keep only the first term in the series of $F_0(\mathbf{k},\bar{\omega},\mathbf{v})$, and $G_s$ can be rewritten as 
\begin{eqnarray}
    &~& G_s(\mathbf{K},\omega) = -\frac{1+p}{2}\frac{k}{\pi}\omega^2_p \label{89} \\
    &~& \quad \quad + 4i\bar{\omega}(p-1) \left(\frac{m}{2\pi\hbar}\right)^3\int_>d^3\mathbf{v}\left(-e^2f'_0\right)v_zF_+(\mathbf{K},\bar{\omega},\mathbf{v}).  \nonumber
\end{eqnarray}
A comparison between this expression and Eq.~(\ref{Gsa}) is displayed in Fig.~\ref{fig:G}; they agree with each other very well, especially for not so big $kv_F/\omega$. The first term of expression (\ref{89}) can be absorbed in $G_b$. It renormalizes the SPW frequencies and renders the latter surface specific, i.e. dependent on the \textit{Fuchs} parameter $p$.  The second term is mostly imaginary and responsible for the aforementioned instability. It is easy to see that $G = 0$ for $p=1$, as expected of specularly reflecting surfaces. Thus, the SRM is not the same as the limit $p=1$, in contrast with its intended meanings. 

With $\Omega$ and $G$, one can obtain $\epsilon_s(k,\omega)$ using the definition, Eq.~(\ref{2a19}). The ensuing expression cannot be further simplified and it is thus not repeated here. 

\subsection{The functions $\chi(\mathbf{K},\mathbf{K}',\omega)$ and $P(\mathbf{K},\omega)$}
To obtain the response functions, let us specify the expressions, (\ref{3.2}) -- (\ref{3.6}) for the electronic distribution to the case where the field represents the probing field. The resulting distribution function is to be called $g_\text{probe}(\mathbf{v},z;\mathbf{k},\omega)$. Substituting this for $g$ in Eq.~(\ref{3.7}), one easily obtains $\mathbf{J}_\text{probe}(z;\mathbf{k},\omega)$ and $S_\text{probe}(\mathbf{K},\omega)$. 

We first establish $\chi(\mathbf{K},\mathbf{K}',\omega)$ by considering the responses in case (ii) described in Sec.~\ref{sec:2b}, to an electrostatic potential. The distribution function follows from Eqs.~(\ref{3.2}) -- (\ref{3.6}). It can be written as $$g_\text{probe} = g_\text{prob} + \Theta(v_z) (p-1) g_\text{pros},$$ where 
\begin{eqnarray}
&~& g_\text{prob}(\mathbf{v},z;\mathbf{k},\omega) = -ef'_0 \varphi(\mathbf{K}',\omega) \times \\ &~& \quad \frac{1}{2} \left[F_+(\mathbf{K}',\bar{\omega},\mathbf{v})\cos(q'z)+iF_-(\mathbf{K}',\bar{\omega},\mathbf{v})\sin(q'z)\right] \nonumber
\end{eqnarray}
and 
\begin{eqnarray} 
g_\text{pros}(\mathbf{v},z;\mathbf{k},\omega) = -\frac{1}{2}ef'_0 \varphi(\mathbf{K}',\omega) F_+(\mathbf{K}',\bar{\omega},\mathbf{v})e^{\frac{\tilde{\omega} z}{v_z}}.
\end{eqnarray}
Now $\mathbf{J}_\text{probe} = \mathbf{J}_\text{prob} + \mathbf{J}_\text{pros}$ accordingly splits, where
\begin{equation}
\mathbf{J}_\text{prob}(z;\mathbf{k},\omega) = \left(\frac{m}{2\pi\hbar}\right)^3\int d^3\mathbf{v} e\mathbf{v} g_\text{prob}(\mathbf{v},z;\mathbf{k},\omega)
\end{equation}
and
\begin{equation}
\mathbf{J}_\text{pros}(z;\mathbf{k},\omega) = (p-1) \left(\frac{m}{2\pi\hbar}\right)^3\int_>d^3\mathbf{v} e\mathbf{v} g_\text{pros}(\mathbf{v},z;\mathbf{k},\omega).
\end{equation}
By the fact that $F_+$ is an even function of $v_z$, one concludes $$J_\text{prob,z}(0;\mathbf{k},\omega) \equiv 0.$$ Straightforward manipulations show that
\begin{equation}
\int^\infty_0 dz \cos(qz) \nabla\cdot\mathbf{J}_\text{prob}(z) = - \varphi(\mathbf{K}',\omega) \frac{K^2\Omega^2(K,\omega)}{8 i\bar{\omega}} ~ \delta(q-q'). 
\end{equation}
Similarly, we have
\begin{eqnarray}
&~&J_\text{pros,z}(0;\mathbf{k},\omega) + \int^\infty_0dz \cos(qz) \nabla\cdot\mathbf{J}_\text{pros}(z;\mathbf{k},\omega) \nonumber\\ 
&=& \frac{1-p}{4}\varphi(\mathbf{K}',\omega) \\ &~& \quad \quad \times \left(\frac{m}{2\pi\hbar}\right)^3 \int_> d^3\mathbf{v} (-e^2f'_0)v_zF_+(\mathbf{K},\bar{\omega},\mathbf{v})F_+(\mathbf{K}',\bar{\omega},\mathbf{v}). \nonumber
\end{eqnarray}
With these expressions we can obtain $S_\text{probe}(\mathbf{K},\omega)$ by use of its definition and thence 
\begin{eqnarray}
&~&C(\mathbf{K},\mathbf{K}',\omega) =  - \frac{K^2\Omega^2(K,\omega)}{8} ~ \delta(q-q')  \label{94} \\ &+& \frac{1-p}{4} i\bar{\omega} \left(\frac{m}{2\pi\hbar}\right)^3 \int_> d^3\mathbf{v} (-e^2f'_0)v_zF_+(\mathbf{K},\bar{\omega},\mathbf{v})F_+(\mathbf{K}',\bar{\omega},\mathbf{v}). \nonumber
\end{eqnarray}
Inserting this into Eqs.~(\ref{2a35}) -- (\ref{2a37}), one obtains the semi-classical response function $\chi(\mathbf{K},\mathbf{K}',\omega)$, which can be written in the following form
\begin{equation}
\chi(\mathbf{K},\mathbf{K}',\omega) = \frac{C(\mathbf{K},\mathbf{K}',\omega) + \epsilon^{-1}_s(k,\omega)\bar{C}(\mathbf{K}',\omega)}{\Omega^2(K,\omega) - \bar{\omega}^2}. 
\end{equation}
with $C$ given by Eq.~(\ref{94}), which further gives $\bar{C}$ via (\ref{2a37}). 

\begin{figure*}
\begin{center}
\includegraphics[width=0.95\textwidth]{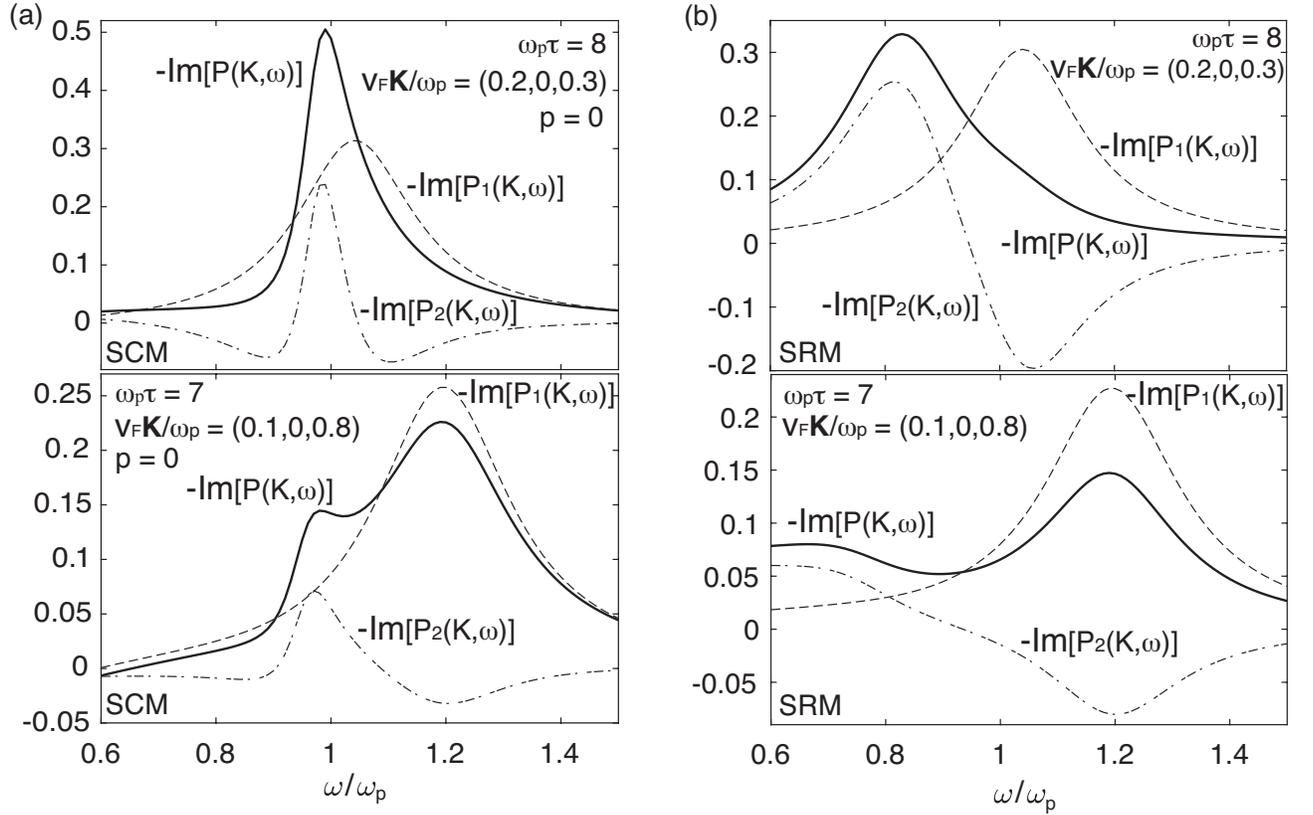}
\end{center}
\caption{Frequency dependence of $P(\mathbf{K},\omega) = P_1(\mathbf{K},\omega) + P_2(\mathbf{K},\omega)$ in (a) the SCM and (b) the SRM at various values of $\mathbf{K}$. \label{fig:Pscm}}
\end{figure*}

The responses to exterior charges are encoded in the function $P(\mathbf{K},\omega)$, which is defined in Sec.~\ref{sec:2b}. One can establish $P$ in a similar fashion as we did with $\chi$, i.e. one could first find the corresponding $g_\text{probe}$ and then uses it to calculate $\mathbf{J}_\text{probe}$ and other quantities including $P$. On the other hand, we can also directly obtain $P(\mathbf{K},\omega)$ from $\chi(\mathbf{K},\mathbf{K}',\omega)$ by means of the relation (\ref{pchi}). For this purpose, it suffices to obtain $B(\mathbf{K},\omega)$ from $C(\mathbf{K},\mathbf{K}',\omega)$ via the relation (\ref{bc}). By the method of contour integral, one can easily show that
\begin{equation}
\frac{1}{\pi} \int^\infty_0 \frac{dq}{K^2} F_+(\mathbf{K},\bar{\omega},\mathbf{v}) = F_0(\mathbf{k},\bar{\omega},\mathbf{v})/k,
\end{equation} 
with which we immediately arrive at
\begin{eqnarray}
&~& B(\mathbf{K},\omega) =  - \frac{\Omega^2(K,\omega)}{4\pi} \label{99} \\ &+& \frac{1-p}{2} i\bar{\omega} \left(\frac{m}{2\pi\hbar}\right)^3 \int_> d^3\mathbf{v} (-e^2f'_0)v_zF_+(\mathbf{K},\bar{\omega},\mathbf{v})F_0(\mathbf{k},\bar{\omega},\mathbf{v})/k. \nonumber
\end{eqnarray}
Here the first term originates from $\nabla\cdot\mathbf{J}_\text{pros}$. Now $P(\mathbf{K},\omega)$ can be directly obtained from these expressions by definition. It can be written as
\begin{equation}
P(\mathbf{K},\omega) = \frac{B(\mathbf{K},\omega) + \epsilon^{-1}_s(k,\omega)\bar{B}(\mathbf{k},\omega)}{\Omega^2(K,\omega) - \bar{\omega}^2}. 
\end{equation}
with $B$ given by Eq.~(\ref{99}), which further gives $\bar{B}$ via (\ref{2a31}). An example of $P$ is plotted in Fig.~\ref{fig:Pscm} (a). At large $K$, the SPWs and VPWs are well separated in frequencies and -Im$[P]$ displays two peaks. 

Setting $p=1$ in the expressions of $B$ and $C$ while neglecting $G_s$, one arrives at the response functions quoted for the SRM, Eqs.~(\ref{srmchi}) and (\ref{srmb}). The as-defined SRM, however, is not identical with the usually adopted SRM, see Ref.~\cite{SI}. In Fig.~\ref{fig:Pscm} a comparison is plotted between the SCM  [panel (a)] and the SRM [panel (b)]. 
 
\section{discussions}
\label{sec:6}
We have developed a general dynamical response theory for SIMs. This theory is straightforwardly extendable to other bounded systems such as films and spheres. We have applied it to discuss the responses within several dispersive and non-dispersive common electron dynamics models in addition to the less common SCM. Analytical expressions have been obtained of the density-density response function $\chi(\mathbf{K},\mathbf{K}',\omega)$, which is probed in virtually every physical process involving surfaces, examples including particle scattering~\cite{mills1975,tsuei1991,nazarov2016} to be discussed in what follows, the scattering of electromagnetic waves~\cite{landau1958}, photon drag effect~\cite{strait2019}, secondary electron emission process (e.g. Auger process) and ion neutralization process~\cite{monreal2014} as well as energy dissipation of objects (e.g. quantum dots and molecules) in the proximity of surfaces~\cite{charles2014,vagov2016} in addition to quantum forces such as quantum friction and Casimir forces~\cite{echenique2011}. These processes are interesting in themselves and they underpin many spectroscopies vital for studying the electronic and optical properties of solids. Applying the theory to these physical processes should be a fascinating subject of future study. 

Our theory requires neither MBCs nor ABCs, which have been avoided by means of the general macroscopic limit of physical boundaries. The entire issue of ABCs has thence been sidestepped. Introduced over six decades ago and having been adopted in innumerable work, ABCs are widely regarded as superficial without a generic physical basis and should not play any role in a complete theory~\cite{apell1984,ginzberg1966}. Our theory reveals that the density response function is comprised of two parts, one of which is directly associated with the excitation of VPWs while the other occurs purely because of the surface capacitive effects and signifies the excitation of SPWs. The ABCs would make the surface part disappear and they are incompatible with non-dispersive models. Our theory calls for a reappraisal of massive experimental data that have been interpreted on the basis of ABCs. 

We are aware of some other work aiming to solve the problem of ABCs. As mentioned in the beginning section of this paper, the earliest effort perhaps dated back to 1970s based on Ewald-Oseen extinction theorem within the dielectric approximation, which has recently been further developed by Schmidt \textit{et al.}~\cite{schmidt2016,schmidt2018}. Another line was taken in the 1990s by Chen \textit{et al.} using their wave-vector-space method~\cite{chen1993,chen}. In the simplest case of local dielectric models, their approach is actually identical to the present one~\cite{chen}. In the development of dispersive models appropriate for the media of excitons, their method is microscopic rather than macroscopic~\cite{chen1993}, allowing them to derive a set of ABCs for the excitons. In addition, K. Henneberger~\cite{hennenberg1998} introduced a controversial source term to mimick the surface effects, which in our opinion resembles the fictitious charge sheet in the SRM and may be regarded as an implicit type of ABCs. Finally, a few years ago~\cite{vaman2009} M. Apostol and G. Vaman also proposed a method that bypasses the ABCs. These authors based their scheme on the concept of a displacement field that is exclusive to the HDM, which is the only model under their consideration. A generalization of their model may be possible if the displacement field is replaced by a more general concept such as the polarization field. As far as the HDM is concerned, their scheme is similar to the present theory.

In the rest of this section, we employ the theory to evaluate the dynamical structure factor, which plays an important role in particle scattering with metal surfaces and in EELS, and the spatial distribution of charges induced by a charged particle grazing over a metal surface. The main purpose here is to differentiate the various dynamics models. We expect the results to be experimentally interesting.  

\begin{figure*}
\begin{center}
\includegraphics[width=0.95\textwidth]{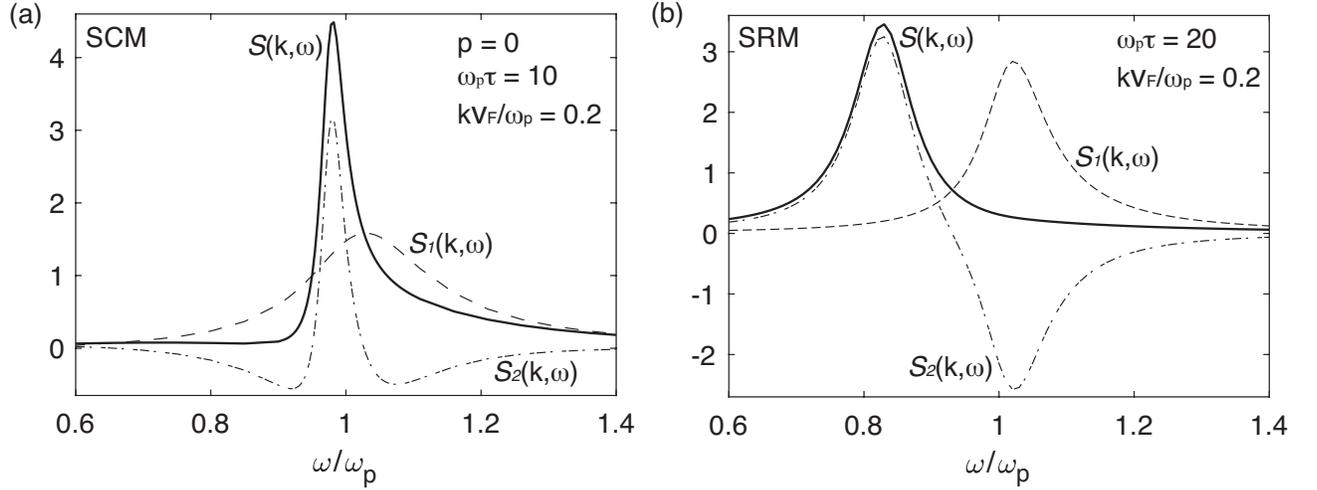}
\end{center}
\caption{The dynamical structure factor $\mathcal{S}(k,\omega) = \mathcal{S}_1(k,\omega) + \mathcal{S}_2(k,\omega)$ in the dipole approximation according to (a) the SCM and (b) the SRM. Only the SPW peak is seen in $\mathcal{S}(k,\omega)$. The peak in the SCM is significantly sharper than in the SRM, even though in the latter a bigger value of $\tau$ has been used. The curve by the HDM -- not shown -- is only slightly different from the SRM curve for the same parameters.\label{fig:chipwa}}
\end{figure*}

\subsection{Dynamical structure factor and SPW peak narrowing}
$\chi(\mathbf{K},\mathbf{K}',\omega)$ is one of the most fundamental quantities for characterizing the responses of a bounded system and pivotal in the interpretation of a variety of experiments. A systematic analysis of its properties being reserved for a separate publication, here we briefly discuss its use in the study of charged particles (e.g. electrons) reflected off a metal surface. The quantity of interest here is the dynamical structure factor $\mathcal{S}$, which appears in the differential scattering cross section per unit surface area (DCS) in the following manner~\cite{nazarov2016},
\begin{equation}
\text{DCS} \propto \frac{K_f}{K_i} \frac{Q^2}{k^2} \mathcal{S}(\Delta\mathbf{K},\omega),
\end{equation}
where $\hbar\mathbf{K}_i$ and $\hbar\mathbf{K}_f$ are the incoming and outgoing momenta of the incident particle of charge $Q$, and $\hbar\Delta \mathbf{K} = \hbar(\mathbf{K}_i - \mathbf{K}_f )= \hbar(\mathbf{k},\Delta k)$ is the momentum exchange during the scattering and $\hbar\omega$ denotes the energy exchange. 

In the so-called dipole approximation~\cite{mills1975}, the particles are assumed to penetrate negligibly into the metal and one has
\begin{equation}
\mathcal{S}(k,\omega) = - \frac{2}{\pi^2} \text{Im}\left[\int^\infty_0 \frac{dq}{K^2} ~ P(\mathbf{K},\omega)\right]. \label{dp}
\end{equation}
Here we have suppressed the dependence of $\mathcal{S}$ on $\Delta k$. In this approximation, it is $P$ that is directly probed rather than the full spectrum of $\chi$.

In Fig.~\ref{fig:chipwa} is exhibited an example of $\mathcal{S}(k,\omega)$, where the left panel is according to the SCM while the right panel to the SRM. The result for the HDM differs only slightly from that for the SRM. In the plots, we have made the decomposition that $\mathcal{S} = \mathcal{S}_1 + \mathcal{S}_2$, where $\mathcal{S}_{1,2}$ are defined via Eq.~(\ref{dp}) with $P$ replaced with $P_{1,2}$; see Sec.~\ref{sec:2}. Only the SPW peak is seeable in $\mathcal{S}(k,\omega)$. This peak is asymmetric in the SCM whereas symmetric in other models -- a result of symmetry breaking effects in $G_s$, which strongly modify the shape of $\mathcal{S}_2(k,\omega)$. As seen in the figure, $\mathcal{S}_1$ has almost the same shape in the SCM as in the SRM, while in the SCM $\mathcal{S}_2$ has a much sharper peak that is far closer to the peak in $\mathcal{S}_1$. At small $k$ this asymmetry becomes less pronounced and eventually disappears.  

As another consequence of the symmetry breaking effects, the width of the SPW peak appears much smaller in the SCM than in other models. It is even much smaller than $1/\tau$, a scenario inexplicable by the conventional wisdom~\cite{rocca1990,khurgin2015}. According to the latter, it can by no means become short of $1/\tau$. This peak narrowing has practical implications for plasmonics and nano photonics, as discussed in recent papers~\cite{deng2018,deng2017a,deng2017b,deng2017c} and briefly recapitulated in Ref.~\cite{SI}. 

The SPW peak width can in principle be made as small as desirable due to a criticality in the system. The criticality can be disclosed in $\mathcal{S}(k,\omega)$. For stable systems, $\mathcal{S}$ must stay positive-definite conforming to the fluctuation-dissipation theorem. For a system containing an instability, however, $\mathcal{S}$ crosses zero at the corresponding critical point to assume unphysical negative values~\cite{saarela1984}. Back to the present case, we note that $\mathcal{S}$ contains two parts $\mathcal{S}_1$ and $\mathcal{S}_2$ canceling each other, as seen in Fig~\ref{fig:chipwa} (a). As shown in Ref.~\cite{SI}, upon decreasing $1/\tau$, $\epsilon_s(k,\omega)$ can be made to vanish and hence $\mathcal{S}_2$ can be made singular around the SPW pole whereas $\mathcal{S}_1$ is dominated by Landau damping via the VPW pole and much less affected. As a result, there exist a critical value of $\tau$, across which $\mathcal{S}$ changes sign from positive to negative near the SPW pole, thereby signifying an instability of the system. At the critical point, SPWs are lossless, as should be for any critical phenomena. In Ref.~\cite{deng2018}, we have put forth a proposal on how to realize this instability. The nature of this criticality is currently under investigation within a quantum mechanical theory. 

To gain some insights into the narrowing of the SPW peak, let us examine the limit of small $k$. It should be cautioned that at very small $k$ retardation effects may play a role and our theory needs to be modified; see Ref.~\cite{SI,deng2015} for discussions on this matter. For very small $k$, we note that $k/K^2 \approx \pi \delta(q)$. Using this, we find $$\mathcal{S}(k,\omega) \approx - \frac{1}{\pi k} \text{Im}\left[P(\mathbf{K}_0,\omega)\right],$$ where $\mathbf{K}_0 = (\mathbf{k},0)$. With the same strategy, we find
\begin{eqnarray}
\bar{B}(k,\omega) &=& \frac{\pi}{2k}\frac{G(\mathbf{K}_0,\omega)B(\mathbf{K}_0,\omega)}{\omega^2_p-\bar{\omega}^2}, \\ 
\epsilon_s(k,\omega) &=& 1 - \frac{\pi}{2k} \frac{G(\mathbf{K}_0,\omega)}{\omega^2_p - \bar{\omega}^2}. 
\end{eqnarray}
Here we have used $\Omega \approx \omega_p$ for small $k$. Expressions of $G$ and $B$ can similarly be found for small $k$. They are given by
\begin{eqnarray}
G(\mathbf{K}_0,\omega) &\approx& \frac{k\omega^2_p}{\pi} \frac{1-p}{2} \left(1 - i\frac{3kv_F}{\bar{\omega}}\right), \\
B(\mathbf{K}_0,\omega) &\approx& -\frac{\omega^2_p}{4\pi} \left(1 + i\frac{3(1-p)}{8}\frac{kv_F}{\bar{\omega}}\right). 
\end{eqnarray}
Combining the above expressions, we obtain
\begin{equation}
P(\mathbf{K}_0,\omega) \approx \frac{B(\mathbf{K}_0,\omega)}{\omega^2_p -\bar{\omega}^2} \frac{1}{\epsilon_s(k,\omega)} = \frac{B(\mathbf{K}_0,\omega)}{\omega^2_s\left(1+i\frac{3(1-p)}{3+p}\frac{kv_F}{\bar{\omega}}\right) - \bar{\omega}^2}. \label{113} 
\end{equation}
Here $\omega_s = \sqrt{\frac{3+p}{4}}\omega_p$. A little more manipulation shows that
\begin{equation}
P(\mathbf{K}_0,\omega) \approx \frac{B(\mathbf{K}_0,\omega)/\left(1+i\frac{3(1-p)}{3+p}\frac{kv_F}{\bar{\omega}}\right)}{\omega^2_s - (\omega+i\gamma)^2}, ~ \gamma = \frac{1}{\tau} - \gamma_0, \label{114}
\end{equation}
where $\gamma_0 = \frac{3(1-p)}{2(3+p)}kv_F$. This expression shows that the effective collision rate $\gamma$ is reduced relative to its bare value $\tau^{-1}$ by an amount of $\gamma_0$. This reduction occurs solely because of the imaginary part of $G_s$, which is absent in other models than the SCM. $P(\mathbf{K}_0,\omega)$ displays a peak at $\omega_s$ with width $\gamma$, which represents the excitation of SPWs. 

As expected, both $\omega_s$ and $\gamma$ depend on surface roughness via the \textit{Fuchs} parameter $p$. Such dependence is absent from other models than the SCM. A detection (an absence) of this dependence would constitute a strong evidence in support of (against) the SCM. Experimentally, it has been demonstrated that $p$ can be widely tuned in some materials such as copper~\cite{cu1,cu2}.  

The long-wavelength SPW frequency in the SCM is $\omega_s \approx 0.87\omega_p$ for diffusely scattering surfaces, which is considerably higher than $0.71\omega_p$ obtained with other models. On the basis of a specific microscopic model within random-phase approximation, Feibelman argued that the SPW frequency should take on the latter value regardless of the microscopic electron density profile near the surface~\cite{feibelman1971}. The solution he found with frequency $0.71\omega_p$ has a constant electrostatic potential and is hence empty of charges, which falls in the category of false solutions mistakenly assigned as standing for SPWs~\cite{deng2018}. To discriminate between these two values, a main difficulty lurks in the determination of $\omega_p$. Let us take Al for the sake of illustration. Nominal charge counting gives $15$eV for $\hbar\omega_p$ in this metal, whereas first principles computation~\cite{chang1994} yields $12.6$eV. Now that the measured SPW frequency~\cite{powell1960} is $10.7$eV in Al, the former would come in favor of $0.71\omega_p$ while the latter of $0.87\omega_p$. This example calls for more effort to be invested in clarifying this issue in the future. 
 
The dipole approximation, despite its widespread use, is incapable of satisfactorily reproducing the experimental observations. In this approximation, $\mathcal{S}(k,\omega)$ displays only the SPW peak, though an additional broad peak due to VPWs has been seen in numerous scattering experiments~\cite{tsuei1991,nazarov2016}. Several proposals have been evoked to address the discrepancy~\cite{nazarov2016}. We shall address this issue comprehensively elsewhere. In the rest of this section, we discuss the issue in terms of the induced charges. 

\subsection{Charges induced by a grazing particle}
For simplicity, let us consider a particle of unit charge grazing over a metal surface at distance $z_0$ and constant velocity $\mathbf{V} = (V,0,0)$, as shown in Fig.~\ref{fig:f1} (a). The associated charge density is given by $\rho_\text{probe}(\mathbf{x},t) = \delta^3(\mathbf{x}-\mathbf{V}t),$ or equivalently $$\rho_\text{probe}(z;\mathbf{k},\omega) = (2\pi/\sqrt{A})\delta(z+z_0)\delta(\omega-k_xV).$$ It follows that $$\xi(\mathbf{k},\omega) = (4\pi^2/\sqrt{A})e^{-kz_0}\delta(\omega-k_xV).$$ The induced charge density is given by
\begin{eqnarray}
\rho(\mathbf{x},t) = \sum_\mathbf{k}\frac{e^{i\mathbf{k}\cdot\mathbf{r}}}{\sqrt{A}}\frac{2}{\pi}\int^\infty_0dq \cos(qz) \int^\infty_{-\infty}\frac{d\omega}{2\pi}\rho(\mathbf{K},\omega)e^{-i\omega t}, \nonumber
\end{eqnarray}
which upon using the results in Sec.~\ref{sec:2b} becomes
\begin{equation}
\rho(\mathbf{x},t) = \int d^2\mathbf{k} e^{i\mathbf{k}\cdot\mathbf{r}(t)} e^{-kz_0} \frac{2}{\pi}\int^\infty_0dq \cos(qz) P(\mathbf{K},k_xV). \label{115}
\end{equation}
where the sum over $\mathbf{k}$ has been converted into an integral and $\mathbf{r}(t) = \mathbf{r}-\mathbf{V}_{\parallel}t$ with $\mathbf{V}_\parallel=(V,0)$. Without loss of generality, $t=0$ is taken in all numerical plots. With the expressions of $P(\mathbf{K},\omega)$ obtained in previous sections, $\rho(\mathbf{x},t)$ can be evaluated. It is noted that the factor $e^{-kz_0}$ effectively suppresses the contributions from $k\gg 1/z_0$ to the integral over $\mathbf{k}$ in the expression. For large $z_0$ only components with small $k$ contribute, whereas for small $z_0$ large-$k$ components also contribute.  

In the DM, $P(\mathbf{K},\omega)$ does not depend on $q$ and hence the induced charge density, which we call $\rho_\text{DM}(\mathbf{x},t)$, is completely localized on the surface, i.e. $\rho_\text{DM}(\mathbf{x},t) = 2\rho_s(\mathbf{r},t)\delta(z)$, with the areal density given by
\begin{equation}
\rho_s(\mathbf{r},t) = \frac{1}{2\pi}\int d^2\mathbf{k} e^{i\mathbf{k}\cdot\mathbf{r}(t)-kz_0}\frac{\omega^2_p/2}{(k_xV+i/\tau)^2-\omega^2_p/2}. 
\end{equation} 
In the limit $z_0\gg V\tau$, one may disregard $k_xV$ and $$\rho_s(\mathbf{r},t) \approx -\frac{1}{2\pi}\frac{\omega^2_p}{2\tau^{-2}+\omega^2_p} \int d^2\mathbf{k} e^{i\mathbf{k}\cdot\mathbf{r}(t) - kz_0},$$ which has a circular shape with radius $\sim z_0$. For not so large $z_0$, the distribution is anisotropic around the grazing particle. An example is shown in Fig.~\ref{fig:f1} (b), where the DM is contrasted with the SCM (of the diffuse limit $p=0$) in terms of the planar charge distribution $\rho_\parallel(\mathbf{r},t) = \int dz~\rho(\mathbf{x},t)$, which equals $\rho_s(\mathbf{r},t)$ in the DM. For small $z_0$ (the panels with $z_0\omega_p/v_F = 5$), in both models $\rho_\parallel(\mathbf{r},t)$ is periodic along the $y$-direction but with a smaller wavelength in the former. For moderate $z_0$ (the panels with $z_0\omega_p/v_F = 15$), however, $\rho_\parallel(\mathbf{r},t)$ strongly depends on the model: in the DM it is symmetric about the grazing particle along its motion but in the SCM the charges are more concentrated in front of the particle. 

\begin{figure*}
\begin{center}
\includegraphics[width=0.95\textwidth]{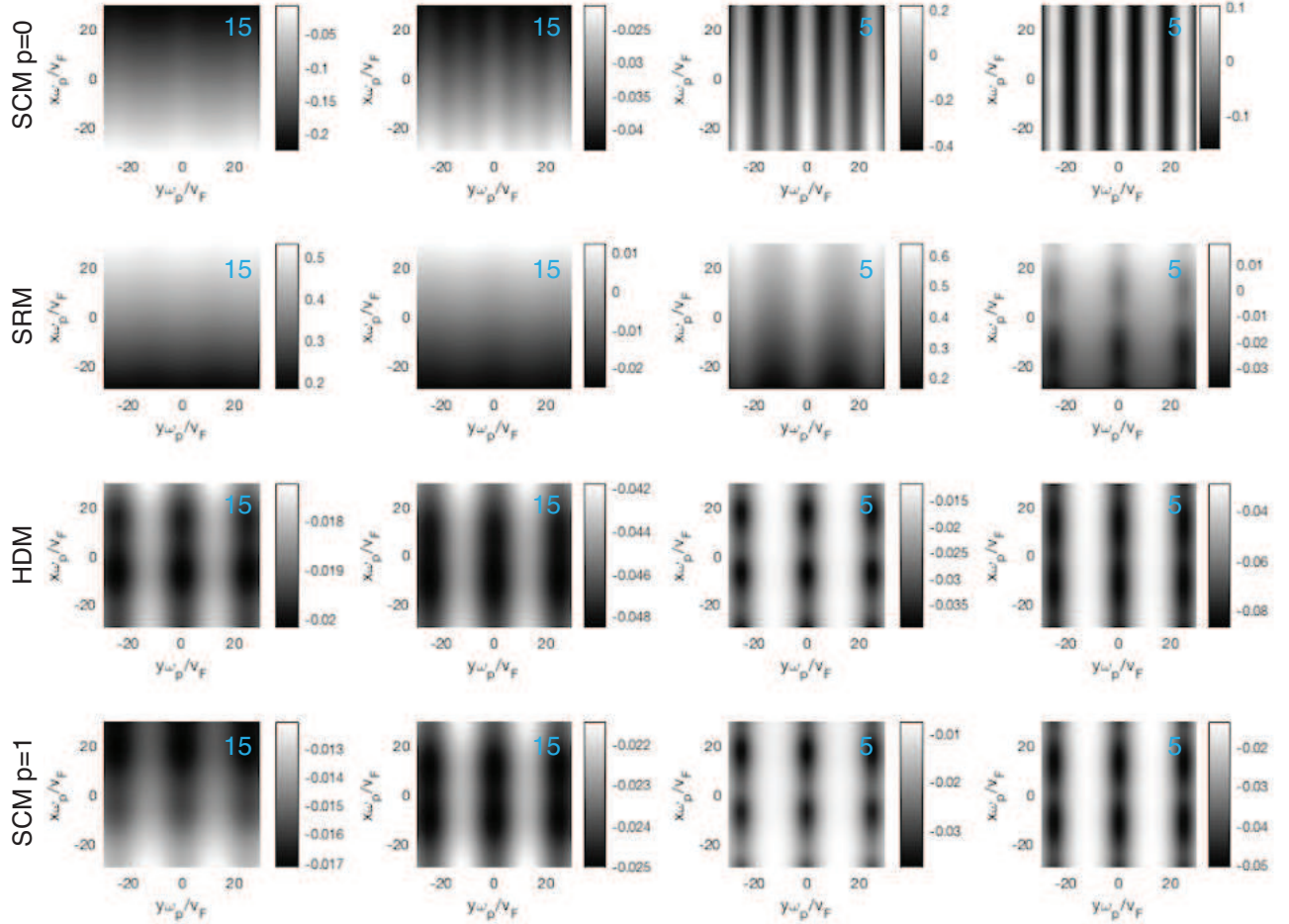}
\end{center}
\caption{Planar distribution of the charges induced by a particle of unit charge grazing over the surface at constant speed $V=10v_F$ and distance $z_0$, see Fig.~\ref{fig:f1} (a). The number at the upper right corner of each panel indicates the value of $z_0\omega_p/v_F$. Within each pair of panels of the same $z_0$ and the same model, the left panel displays $\rho(\mathbf{x}_0,t)$ and the right one displays $\rho_\parallel(\mathbf{r},t) = \int^\infty_0 dz ~ \rho(\mathbf{x},t)$. The particle is located at $(0,0,-z_0)$ for the moment under consideration. Gray scale indicates their values.\label{fig:f7}}
\end{figure*}

The aforementioned symmetry is preserved in the HDM but not in the SRM, as seen in Fig.~\ref{fig:f7}. In this figure, the panels are organized in eight pairs, each pair consisting of two panels in the same model and with the same $z_0$. The left panel in a pair shows $\rho(\mathbf{x}_0,t)$ while the right one shows $\rho_\parallel(\mathbf{r},t)$. For comparison, we have also displayed results for the SCM of the reflection limit $p=1$. For small $z_0$, $\rho_\parallel(\mathbf{r},t)$ exhibits in the SRM, the HDM and the SCM of $p=1$ the same periodic and symmetric pattern as in the DM, though its magnitude strongly depends on the models. For moderate $z_0$, $\rho_\parallel(\mathbf{r},t)$ remains symmetric in the HDM and the SCM of $p=1$ but not so in the SCM of $p=0$ and the SRM. In general, $\rho(\mathbf{x}_0,t)$ varies much more mildly than $\rho_\parallel(\mathbf{r},t)$ along the surface. 

\begin{figure*}
\begin{center}
\includegraphics[width=0.95\textwidth]{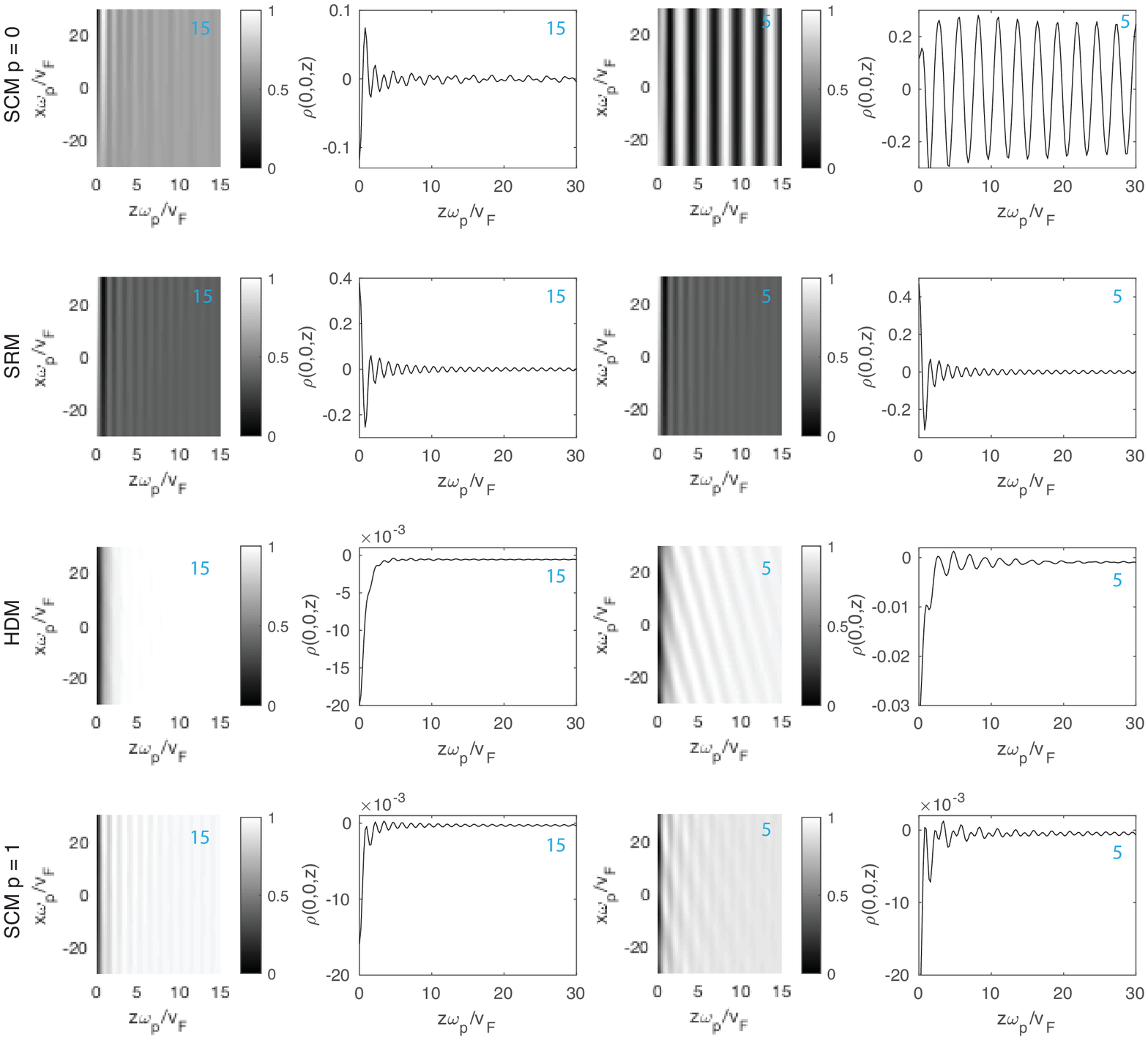}
\end{center}
\caption{Depth distribution of the charges induced by the grazing particle. The number at the upper right corner of each panel indicates the value of $z_0\omega_p/v_F$. Within each pair of panels of the same $z_0$ and the same model, the left panel displays $\rho(\mathbf{x},t)$ in the plane $y=0$ and the right one displays $\rho[(\mathbf{r}_0,z),t]$, where $\mathbf{r}_0 = (0,0)$. Gray scale indicates their values. The particle is located at $(0,0,-z_0)$ for the moment under consideration. \label{fig:f8}}
\end{figure*}

The depth dependence of the induced charge density is illustrated in Fig.~\ref{fig:f8}. Here the panels are also grouped in eight pairs, each consisting of two panels in the same model and with the same value of $z_0$. The left panel in a pair displays the distribution of the induced charges in the plane $y=0$ while the right one displays $\rho[(\mathbf{r}_0,z),t]$ versus $z$, where $\mathbf{r}_0 = (0,0)$. For big $z_0$, in all models $\rho[(\mathbf{r}_0,z),t]$ decays quickly away from the surface, indicating that the charges are strongly concentrated about the surface. For small $z_0$, however, $\rho[(\mathbf{r}_0,z),t]$ oscillates in the SCM of $p=0$. This oscillation stems from symmetry breaking effects encoded in $G_s$ and $B_s$ that are absent from other models, and it is associated with the excitation of VPWs. In the SCM of $p=1$, $P_2(\mathbf{K},\omega)$ vanishes and $P(\mathbf{K},\omega) = (1/4\pi)\Omega^2(\mathbf{K},\omega)/(\bar{\omega}^2-\Omega^2(\mathbf{K},\omega))$. As $\Omega$ varies only slightly with $q$ when $\omega_p\tau$ is not very large, the resulting $\rho(\mathbf{x},t)$ is also largely localized on the surface as seen in this figure, closely resembling that of self-sustained SPWs, whose density is $\rho_\text{SPW}(\mathbf{K}) = \text{const}/\epsilon(K,\omega)$, though only VPWs are excited in the limit of $p=1$. 

The induced charge density profiles are of experimental interest for two reasons. Firstly, they may be directly measured~\cite{huang2014} to discriminate existing models against one another. In particular, the validity of the SCM can be examined. Secondly, the induced charge density is ultimately responsible for the energy losses experienced by the probing particles. Such losses can be measured to benchmark the models. A systematic study of this issue is beyond the scope of the present paper and will be published elsewhere. 

\section{Summary}
\label{sec:7}
In summary, we have presented a general macroscopic theory of electrodynamic response for bounded systems without the use of ABCs and MBCs. The theory yields analytical expressions of the density-density response function and sheds fresh light into its mathematical structure and the physical origin behind it. It provides a physically transparent way of evaluating the function either analytically or numerically. Such transparency is not affordable in existing calculations. A unified view has been rendered of various dispersive and non-dispersive models, including the DM, the HDM, the SRM and the SCM. Some long-standing misconceptions regarding these models have been clarified. 

According to the SCM, an intrinsic instability of the metal is predicted to occur, as may be revealed as a zero of the dynamical structure factor. This instability can be utilized to drastically reduce the energy losses suffered by SPWs that have so far impeded the progress in the field of plasmonics, as suggested in our previous work. 

In contrast with conventional wisdom, we find that a grazing exterior charge can excite volume density waves in a SIM provided the charge is in the vicinity of its surface. We also find that the distribution of induced charges is sensitive to the dynamics model in use. The SCM distinguishes itself from other common models by the inclusion of effects due to translation symmetry breaking and surface roughness. A measurement of the charge distribution may be carried out to examine the validity and limitations of these models. 

While it is explicitly developed for metals, in which electrical currents are carried primarily by conduction electrons, the general theory as developed in Sec.~\ref{sec:2} can be adapted to situations where the currents may be of a different nature, e.g. due to excitons. 

Addressing a fundamental problem in condensed matter physics and surface science, the theory is expected to be useful in a number of areas including chemistry and nuclear instruments design. Applications in particle scattering and light scattering as well as other phenomena such as quantum forces will be explored in the future.  

\textbf{Acknowledgement.} The author is grateful to Dr. T. Philbin for bringing to his notice the recent work in Refs.~\cite{schmidt2016, schmidt2018}. He also thanks J. Pendry and M. Apostol for some useful suggestions.

\end{document}